# Doping-driven topological polaritons in graphene/α-MoO₃ heterostructures

Hai Hu,[1,2†*] Na Chen,[1,2†] Hanchao Teng,[1,2†] Renwen Yu,[3,4*] Yunpeng Qu,[1,2] Jianzhe Sun,[7] Mengfei Xue,[8] Debo Hu,[1,2] Bin Wu,[7] Chi Li,[1,2] Jianing Chen,[8] Mengkun Liu,[9] Zhipei Sun,[10] Yunqi Liu,[7] Peining Li,[6] Shanhui Fan,[4] F. Javier García de Abajo,[3,5*] Qing Dai[1,2*]

1 CAS Key Laboratory of Nanophotonic Materials and Devices, CAS Key Laboratory of Standardization and Measurement for Nanotechnology, CAS Center for Excellence in Nanoscience, National Center for Nanoscience and Technology, Beijing 100190, P. R. China.
2 University of Chinese Academy of Sciences, Beijing 100049, P. R. China.
3 ICFO-Institut de Ciencies Fotoniques, The Barcelona Institute of Science and Technology, 08860 Castelldefels (Barcelona), Spain.
4 Department of Electrical Engineering, Ginzton Laboratory, Stanford University, Stanford, California 94305, United States.
5 ICREA-Institució Catalana de Recerca i Estudis Avançats, Passeig Lluís Companys 23, 08010 Barcelona, Spain.
6 Wuhan National Laboratory for Optoelectronics and School of Optical and Electronic Information, Huazhong University of Science and Technology, Wuhan, P. R. China.
7 Beijing National Laboratory for Molecular Sciences, Key Laboratory of Organic Solids, Institute of Chemistry, Beijing 100190, P. R. China.
8 The Institute of Physics, Chinese Academy of Sciences, P.O. Box 603, Beijing, China.
9 Department of Physics and Astronomy, Stony Brook University, Stony Brook, New York 11794, USA.
10 Department of Electronics and Nanoengineering Aalto University Tietotie 3, FI-02150 Espoo, Finland.

*e-mail: daiq@nanoctr.cn, javier.garciadeabajo@nanophotonics.es, renwen.yu@icloud.com, huh@nanoctr.cn

† These authors contributed equally

## Keywords

**Abstract**

Controlling the charge carrier density provides an efficient way to trigger phase transitions and modulate the optoelectronic properties in natural materials. This approach could be used to induce topological transitions in the optical response of photonic systems. Here, we predict a topological transition in the isofrequency dispersion contours of hybrid polaritons supported by a two-dimensional heterostructure consisting of graphene and α-phase molybdenum trioxide (α-$MoO_3$). By chemically changing the doping level of graphene, we experimentally demonstrate that the contour topology of polariton isofrequency surfaces transforms from open to closed shapes as a result of doping-dependent polariton hybridization. Moreover, by changing the substrate medium for the heterostructure, the dispersion contour can be further engineered into a rather flattened shape at the topological transition, thus supporting tunable polariton canalization and providing the means to locally control the topology. We demonstrate this idea to achieve extremely subwavelength focusing by using a 1.2-µm-wide silica substrate as a negative refraction lens. Our findings open a disruptive approach toward promising on-chip applications in nanoimaging, optical sensing, and manipulation of nanoscale energy transfer.

## Introduction

The control of charge carrier concentration by either electrostatic or chemical means has been widely studied as a way to induce phase transitions, such as structural phase transitions in transition-metal dichalcogenides[1-5], ferromagnetic phase transitions in high-Curie-temperature manganites[6-11], and topological transitions[12-15], with potential application in the development of active electronic phase-change devices[16]. In this context, a recent study shows that a collection of different phases in magic-angle bilayer graphene can be achieved by changing its carrier density[17]. Similar concepts have been theoretically explored in photonics using hyperbolic metamaterials composed of subwavelength structures, such as a periodic array of graphene ribbons[18] or a stack of graphene-dielectric layers[19], where a topological transition in the isofrequency dispersion contour can occur by changing the doping level of graphene. However, these hyperbolic metamaterials rely on a strong anisotropy of the effective permittivity tensor, which is ultimately limited by spatial nonlocal effects that can hinder a practical verification of this concept.

Recently, a twisted stack of α-phase molybdenum trioxide (α-$MoO_3$) slabs has been explored to control the topology of the isofrequency dispersion contour of phonon polaritons (PhPs) through varying the relative twist angle between the two α-$MoO_3$ layers[20-23]. Thanks to the in-plane anisotropy of the permittivity within the Reststrahlen band from 816 $cm^{-1}$ to 976 $cm^{-1}$, the real part of the permittivity is positive along the [001] direction but negative along the [100] direction,[24,25] which makes α-$MoO_3$ a natural hyperbolic material supporting in-plane hyperbolic PhPs[26,27]. The low-loss in-plane hyperbolic PhPs in α-$MoO_3$ thus emerge as an ideal platform to explore further possibilities on doping-driven and electrically tunable topological transitions in photonics.

In this work, we experimentally demonstrate the control of polariton dispersion in a van der Waals (vdW) heterostructures composed of an α-$MoO_3$ film covered by monolayer graphene through changing the doping level of the latter. The polariton dispersion contour varies from hyperbolic (open) to elliptic (closed) when increasing the doping level of graphene, leading to the emergence of a mode dominated by its graphene plasmon polariton (GPP) component propagating along the [001] direction at high doping levels. The nature of the polaritons emerging at high doping in the heterostructure evolves from GPP to PhP when moving from [001] to [100] α-$MoO_3$ crystallographic directions. In addition, when the vdW heterostructure is placed on top of a gold substrate instead of $SiO_2$, a rather flattened dispersion contour can be obtained as the topological transition takes place. As an application, by engineering the substrate, we demonstrate an in-plane subwavelength focusing device based on negative refraction. Our findings open exotic avenues for on-demand control of light flow at the nanoscale.[28-30]

## Results and discussion

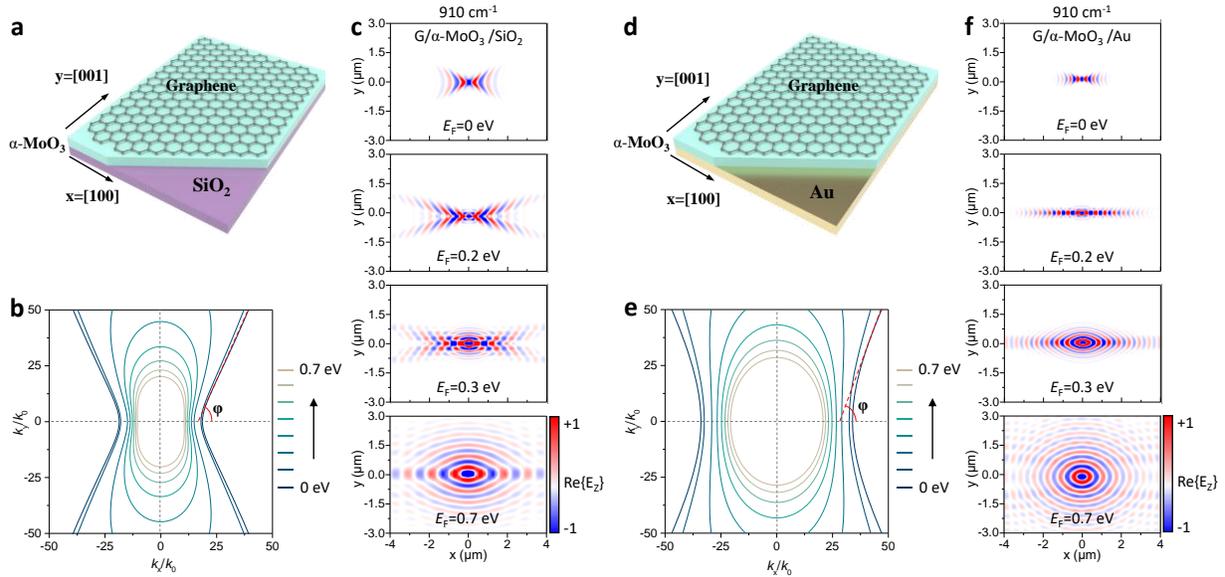

**Figure 1. Topological transition of hybrid polaritons**. **(a, d)** Illustration of the graphene/α-MoO$_3$ vdW heterostructure used in this study, supported on SiO$_2$ (a) and gold (d) substrates, respectively. **(b, e)** Calculated isofrequency dispersion contours of hybrid polaritons on a 300-nm-thick SiO$_2$ substrate (b) and 60-nm-thick gold substrate (e), respectively, at a fixed incident frequency of $\lambda_0$= 910 cm$^{-1}$ for different graphene Fermi energies ranging from 0 to 0.7 eV (in steps of 0.1 eV, see colored labels). The thickness of the α-MoO$_3$ film is 150 nm. **(c, f)** Numerically simulated field distributions (real part of the z out-of-plane component of the electric field, Re $\{E_z\}$, of hybrid polaritons on SiO$_2$ (c) and gold (f) substrates, respectively, for several graphene doping levels, as launched by a dipole placed 100 nm above the origin.

A schematic of our proposed structure is shown in Figure 1, where a 150-nm-thick vdW heterostructure is placed on top of either a SiO$_2$ (Figure 1a) or gold (Figure 1d) substrate. Here, we are particularly interested in the Reststrahlen band II of α-MoO$_3$ from 816 cm$^{-1}$ to 976 cm$^{-1}$, where the permittivity components along the [100], [001], and [010] crystal directions satisfy $\varepsilon_x$< 0, $\varepsilon_y$> 0, and $\varepsilon_z$> 0, respectively (Supplementary Materials Figure S1a). [16,17] As a result, the in-plane PhPs in natural α-MoO$_3$ exhibit a hyperbolic dispersion contour. To illustrate dynamic control of the dispersion contour topology of the in-plane hybrid plasmon-phonon polaritons in our structure, dispersion contours on SiO$_2$ at different graphene Fermi energies ($E_F$) under a fixed representative incident wavelength $\lambda_0$=10.99 μm (frequency 910 cm$^{-1}$) are presented in Figure 1b. As the graphene Fermi energy increases from 0 to 0.7 eV, the opening-angle φ of the hyperbolic sectors gradually increases due to a change of the PhP wavelength when varying the dielectric environment, and finally the dispersion contour changes its character from hyperbolic (open) to elliptic (closed "8") shape. Note that the Fermi energy at which this topological transition occurs is conditioned by the appearance of well-defined graphene plasmons along the [001] direction at $\lambda_0$ when increasing its doping level. When the substrate is changed to gold, we find more flattened dispersion contours, as shown in Figure 1e, due to the screening provided by the gold substrate. A tunable polariton canalization and diffractionless propagation are thus expected.

Numerically simulated field distributions of hybrid polaritons for different graphene Fermi energies are shown in Figure 1c, f. At $E_F = 0$, the polaritons of the heterostructure on a SiO$_2$ substrate exhibit a hyperbolic wavefront, similar to that of PhPs in α-MoO$_3$. In contrast, the wavelength of polaritons on a gold substrate is highly compressed, while their opening-angle is increased. When increasing the doping level to $E_F = 0.2$ eV, a topological transition arises. The wavelength of hybrid polaritons is increased compared with the undoped case, and we still find a hyperbolic wavefront along the $x$ direction on the SiO$_2$ substrate. More interestingly, for hybrid polaritons on the gold substrate (Figure 1f), highly collimated and directive hybrid polaritons propagating along the $x$ direction are observed as a result of a rather flattened dispersion contour. At $E_F = 0.3$ eV, a hyperbolic wavefront of hybrid polaritons on the SiO$_2$ along the $x$ direction can still be observed, but another mode with an elliptic wavefront propagating along the $y$ direction also appears. As for the hybrid polaritons on the gold substrate, the wavefront is dominated by a fine crescent shape along the $x$ direction. At a very high doping level $E_F = 0.7$ eV, the dispersion contours for hybrid polaritons on both SiO$_2$ and gold substrates hold an elliptic-like shape (Figure 1b, e). As a consequence, we can find modes propagating anisotropically in the $x$-$y$ plane. More details about tuning the topological transition by varying the graphene doping level are shown in Figures S2-S4 in the Supplementary Materials.

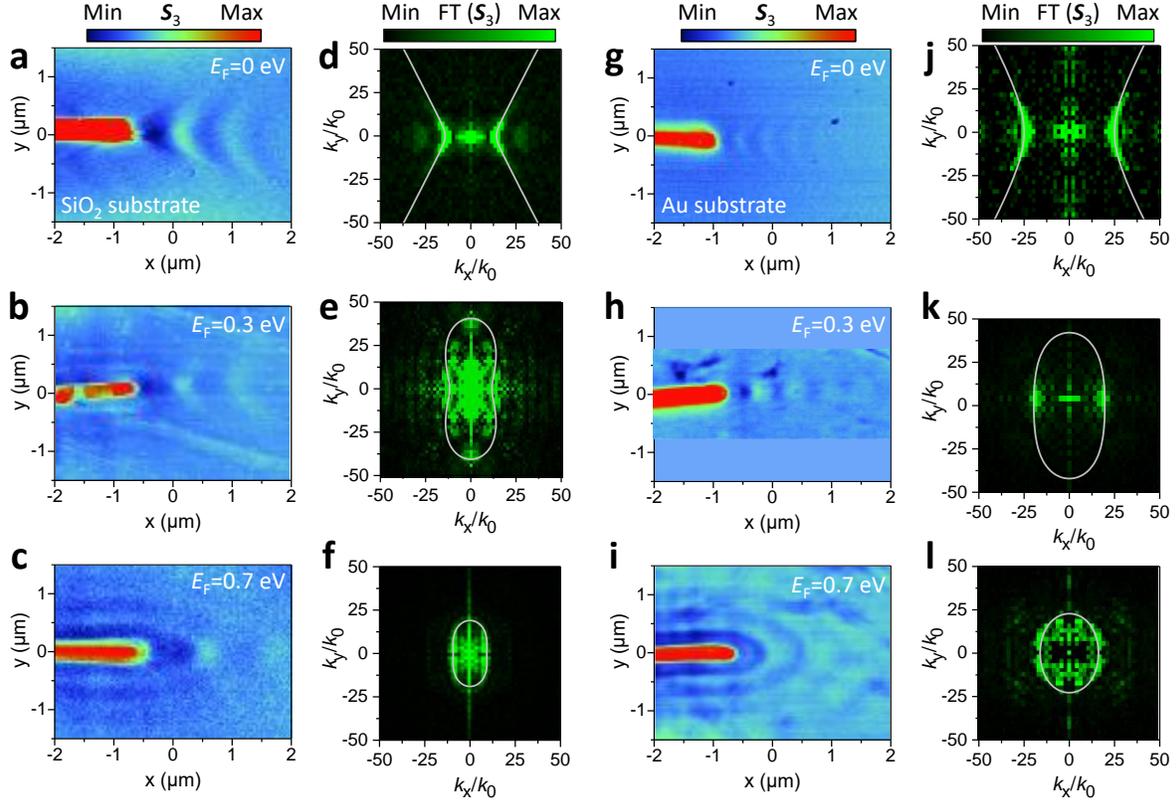

**Figure 2. Topological transition of hybrid polaritons revealed by nano-imaging. (a-c)** Experimentally measured polariton near-field distributions with graphene doping $E_F$=0, 0.3, and 0.7 eV. The polaritons are launched by a gold antenna. The α-MoO$_3$ film is placed on top of a 300-nm-thick SiO$_2$/500-μm-thick Si substrate. **(d-f)** Absolute value of the spatial Fourier transforms of the experimental near-field images shown in (a-c), which reveal the isofrequency contours of hybrid polaritons. The grey curves represent the calculated isofrequency contours. **(g-i)** Experimental measurements similar to those shown in (a-c), but with the substrate changed to 60-nm-thick Au/500-μm-thick Si. The canalized wavefronts are measured at a graphene Fermi energy close to the value for which the topological transition occurs (h), showing deep subwavelength and diffractionless polariton propagation. **(j-l)** Absolute value of the Fourier transforms of the experimental near-field images in (g-i) with grey curves showing the calculated isofrequency contours. The α-MoO$_3$ thickness is 140 nm in all panels. The incident light wavelength is fixed at $\lambda_0$ = 11.11 μm (900 cm$^{-1}$).

Infrared nano-imaging visualizing the propagating polaritons in the graphene/α-MoO$_3$ heterostructures (Figure S5) unambiguously supports the above theoretical predictions, as shown in Figures 2a-c and 2g-i. Upon p-polarized infrared light illumination, the resonant gold antenna efficiently launches hybrid polaritons (Figure S6). While scanning the sample, the real part of the tip-scattered light electric field (Re $\{E_S\}$) is recorded simultaneously with topography, making it possible to directly map the vertical near field components of the hybrid polariton wavefronts launched by the antenna. Figures S7 and S8 show polariton interferometry images formed in our sample and measured at different incidence frequencies and graphene Fermi energies. In the sample, polariton fringes of two distinct periodicities ($\lambda_p$ and $\lambda_p/2$) are observed near the sample edge, the graphene boundary, and the edge of the gold antenna. The $\lambda_p/2$-period fringes can be assigned to the tip-launched polaritons reflected at the α-MoO₃ edge or the graphene boundary. In addition, the $\lambda_p$-period fringes can be associated with antenna or edge launched polaritons propagating to the tip and being scattered to the detector.

To visualize the polariton wavefronts, we image antenna-launched polaritons in the heterostructure at several intermediate graphene Fermi energies ranging from $E_F$=0 to 0.7 eV (Figure S9). We first investigate hybrid polaritons in an undoped sample with $E_F$=0 eV, revealing a precise hyperbolic wavefront (Figures 2a, g), for both SiO₂ and gold substrates, which is basically consistent with previously observed results in a single slab of α-MoO₃ as a result of the hyperbolic dispersion contour. Next, we examine the optical response of a sample with a relatively high doping level $E_F$=0.3 eV on a SiO₂ substrate at the same incidence frequency (Figure 2b). The measured wavefronts remain hyperbolic along the $x$ direction, while fringes around the antenna appear towards the $y$ direction, indicating that the dispersion contour has transformed into a closed shape, as shown in Figures 1b and S2. As for the sample on a gold substrate (Figure 2h), we find a nearly flat wavefront for $E_F$=0.3 eV, indicating the occurrence of a topological transition in the dispersion contour along the $x$ direction. When the doping level is further increased to $E_F$=0.7 eV, only elliptical wavefronts are observed (Figures 2c, i) for samples on both SiO₂ and gold substrates, denoting a rather rounded anisotropic dispersion contour. The corresponding Fourier transforms of the experimental near-field images (Figures 2d-f and 2j-l) and simulated near-field distributions (Supplementary Materials Figures S3-4) have further confirmed the transformation of the dispersion contour with increasing doping level of graphene. Noticeably, our extracted experimental polariton wave vectors $k = 2\pi/\lambda_p$ (dotted symbols in Figure S10) match quite well with the calculated dispersion diagrams in all cases.

These observations provide experimental evidence for our theoretical prediction (Figures 1 and S2) of doping-induced topological transition and extreme dispersion engineering controlled by the graphene Fermi energy. Similar topological transitions have also been observed at other incidence frequencies (Figures S7 and S8) and different thicknesses of α-MoO₃ (Figures S11-S12). Low levels of disorder or minor imperfections in the heterostructure should not substantially affect the control capability (Figure S13). Therefore, controlling the dispersion contour through varying the graphene doping level shows inherent robustness. On the other hand, the thickness of α-MoO3 determines the influence from the dielectric environment (the substrate), and its thickness should not exceed the skin depth of hybrid polaritons (Figure S14)

Close to $E_F$=0.3 eV, where the dispersion contour is rather flat as shown in Figures 2h, the propagation of hybrid polaritons appears to be firmly guided along the *x* direction, yielding a highly directive and diffractionless character. Furthermore, this type of polariton canalization can be found over a wide range of frequencies and thicknesses of α-MoO$_3$ (Figure S11) due to the inherent robustness of the topological transition. The line profiles (vertical cuts along the *y* direction) across the amplitude of the canalization mode give a full-width-at-half-maximum (FWHM) of around 115 ± 5 nm (≈ $\lambda_0$/95; where $\lambda_0$ is the free-space wavelength), as shown in Figure S12.

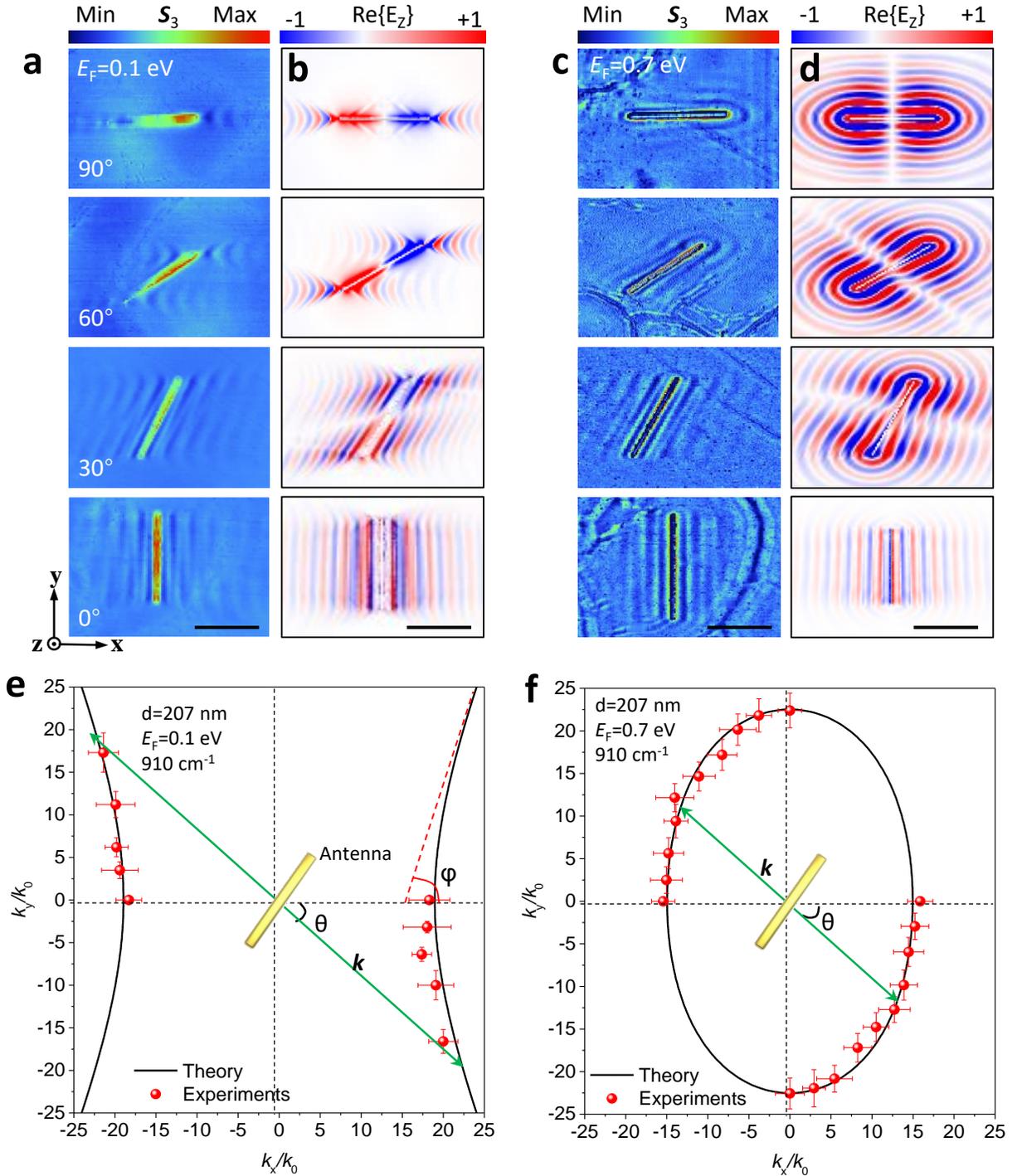

**Figure 3. Antenna-tailored launching of hybrid polaritons.** (a) Experimentally measured near-field amplitude images of hybrid polaritons launched by the gold antennas with θ in the 0° to 90° range (Figure S15). The Fermi energy of graphene is set to $E_F$=0.1 eV and the light frequency is 910 cm$^{-1}$. (b) Corresponding numerically simulated field distributions (Re {$E_z$} evaluated 20 nm above the surface of the heterostructure) to be compared with the measured results shown in (a). (c) Same experimental measurements as in (a), but with a graphene Fermi energy $E_F$=0.7 eV. (d)

Near-field numerical simulations corresponding to (c). (**e**) Isofrequency dispersion contour extracted from the experimental data in (a) (red symbols), compared with the calculated hyperbolic dispersion contour (black solid curve) for an opening angle φ. The green arrow illustrates the direction of the exciting polariton wave vector **k** perpendicular to the long axis of the gold antenna. (**f**) Same as in (e), but extracted from the experimental data in (c). The α-MoO$_3$ thickness is 207 nm in all panels. The scale bars in (a-d) are 2 μm.

By rotating the long axis of the antenna by an angle $\theta$ with respect to the $y$ direction, one can selectively launch and manipulate hybrid polaritons with different in-plane wave vectors. The launching contribution from our antenna can be decomposed into four parts: two endpoints with resonant dipoles and two parallel edges with discrete point dipoles (similar to a line dipole). Note that the incident light is polarized along the $x$ direction.

For the sample with a low doping level ($E_F$=0.1 eV, Figure 3a), at $\theta = 0°$, the field pattern of the exciting hybrid polaritons exhibits vertical fringes parallel to the long axis of the antenna, dominated by the line dipole contributions generated by the edges. Note that the two endpoint dipoles are not well excited when $\theta = 0°$ because the polarization direction of the incident light is not aligned with the long axis of the antenna. One can extract the polariton wave vector from the fringes parallel to the long axis of the antenna. When $\theta$ increases from 0° to 90°, the wavefronts produced by the two endpoints gradually appear and interact with those produced by the edges. In the regime with rotating angles $\theta \leq 40°$, the distance between adjacent fringes parallel to the antenna edge is reduced from 590 to 380 nm, from which one can obtain the wave vector **k** perpendicular to the antenna. The extracted wave vector is shown as red symbols in Figure 3e, matching quite well with the calculated dispersion contour (solid curves; more details about the extraction analysis are offered in Figure S16 of the Supplementary Materials. When $\theta \geq 60°$ ($\approx$φ, the opening-angle indicated in Figure 3e), there are no fringes parallel to the edge of the antenna because polariton propagation is prohibited along that direction judging from the dispersion contour, and the field pattern is dominated by the hyperbolic wavefronts produced by two endpoints of the antenna. The simulated field patterns (Figure 3b) corroborate our experimental observations for various rotating angles.

With the sample at a high doping level ($E_F$=0.7 eV, Figure 3c), the antenna can generate polaritons propagating in all directions within the $x$-$y$ plane when the polarization direction of the incident light is along the long axis of the antenna. Due to the in-plane anisotropy, the excited field patterns are therefore different for various rotating angles ranging from 0° to 90°. The simulated field patterns (Figure 3d) again agree well with our experimental observations. One can still extract the polariton wave vector from the measured fringes perpendicular to the antenna, shown as red symbols in Figure 3f, which also match quite well with the calculated dispersion contour (solid curves). Note that, at $E_F=0.3$ eV, the field patterns of hybrid polaritons launched by antennas with different rotating angles (from 0 to 45°) are all strictly along the *x* direction (Figure S17) due to the flattening of the dispersion contour, which leads to a directional canalization at this graphene Fermi energy.

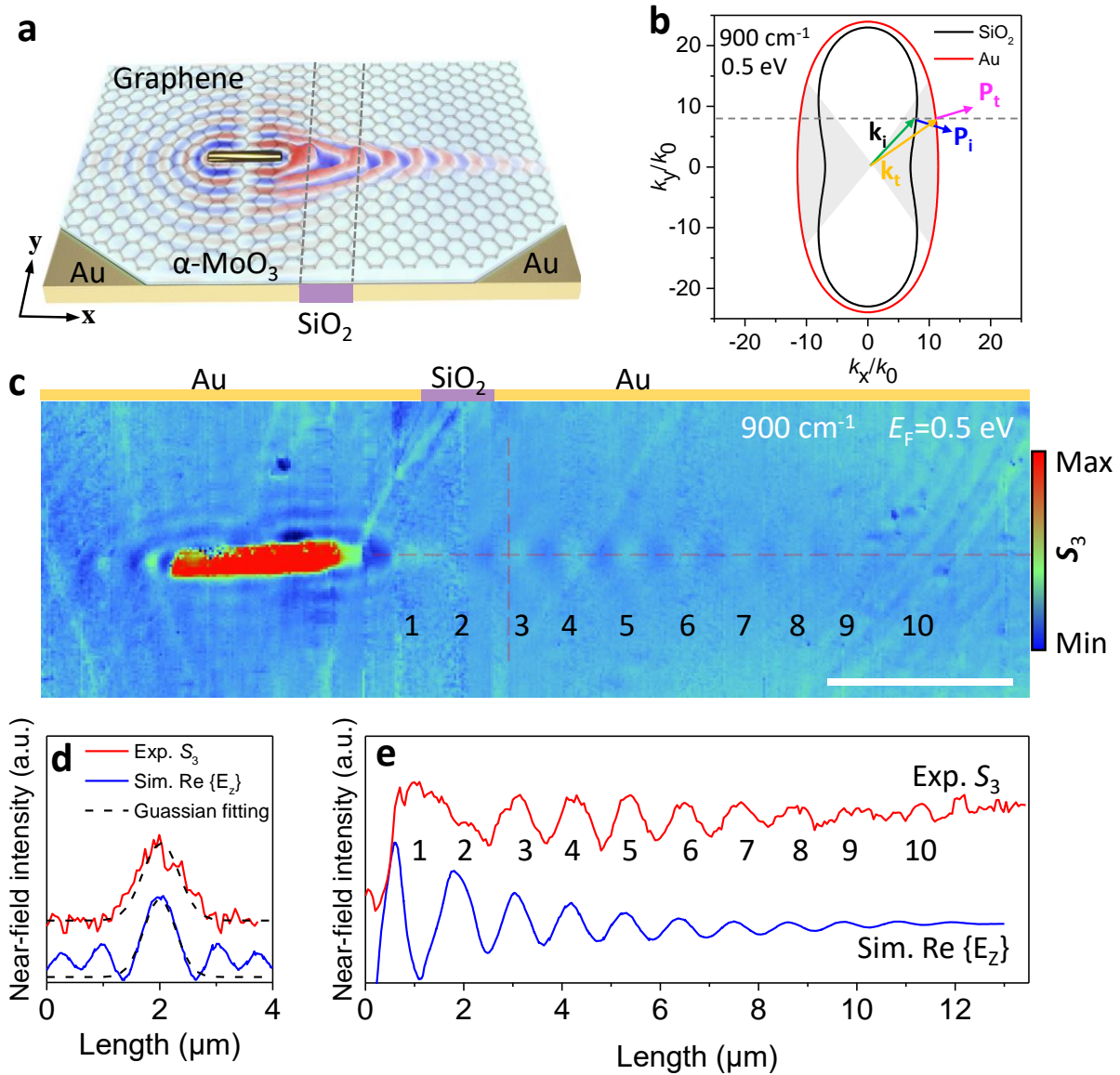

**Figure 4. Partial focusing of hybrid polaritons by substrate engineering. (a)** Schematics of the design, where the heterostructure lies on top of a substrate composed of an Au-SiO$_2$-Au in-plane sandwich structure. **(b)** Isofrequency dispersion contours of hybrid polaritons for Au (red) and SiO$_2$ (black) substrates at $\lambda_0 = 900$ cm$^{-1}$ (11.11 μm). The shaded areas highlight convex and concave dispersion contours in the region around the *x* axis on the gold and SiO$_2$ substrates, respectively. With a wave vector inside the shaded area, negative refraction can happen at the Au-SiO$_2$ interface when the polaritons on the gold substrate propagate towards that interface. The scheme for negative refraction is illustrated by further showing the incident wave vector **k$_i$** and Poynting vector **P$_i$** together with the resulted transmitted **k$_t$** and **P$_t$**. (**c**) Experimentally measured partial focusing of hybrid polaritons in the system shown in panel (a). The central SiO$_2$ slab is 1.2 µm wide and works as an in-plane flat lens. The antenna is located 0.9 µm away from the left Au-SiO$_2$ interface. The scale bar indicates 3 µm. (**d**) Near-field profiles are taken at the positions marked by red (c) and blue (Figure S22) vertical dashed lines. The black dashed curves are Gaussian fittings. (**e**) Near-field profiles at the positions marked by the red (c) and blue (Figure S22) horizontal dashed lines. The graphene is doped to $E_F$=0.5 eV and the α-MoO$_3$ thickness is 240 nm in both experiments and simulations.

From the results displayed in Figures 1 and 2, one can see that the dispersion contour of the hybrid polaritons can be modified significantly by changing the substrate. In particular, at high doping levels, the dispersion contour is convex in the region around the *x* axis on the gold substrate, whereas it is concave on the SiO$_2$ substrate (see the shaded area in Figure 4b). Therefore, control of the dielectric environment of the heterostructure can provide alternative possibilities to manipulate the in-plane propagation of hybrid polaritons.

The proposed design is illustrated in Figure 4a, where the heterostructure lies on top of a substrate composed of an Au-SiO$_2$-Au in-plane sandwich structure. This substrate is used to locally engineer the isofrequency dispersion contour (Figure 4b). The central SiO$_2$ slab, with a width of 1.2 µm, serves as an in-plane flat lens to partially focus the incident polaritons (with a wave vector **k$_i$** and a Poynting vector **P$_i$** along the normal of the contour in Figure 4b) generated by an antenna on top of the left gold substrate. When hybrid polaritons cross the boundary between gold and SiO$_2$ substrates, due to the change in the detailed shape of the dispersion contour, negative refraction can occur at the boundary with the *y* component of the wave vector conserved whereas the sign of the *y* component of the transmitted Poynting vector **P$_t$** is opposite that of the incident **P$_i$**, as illustrated in Figure 4b, which would lead to a partial focusing of the polaritons. Figure S18 shows the evolution of negative refraction of hybrid polaritons at different Fermi energies of graphene. The measured antenna-launched polariton near-field distributions for the heterostructure on gold and SiO$_2$ substrates are shown in Figure S19, respectively. On the gold substrate (Figure S19a), only elliptical wavefronts are observed around the antenna, denoting a convex dispersion contour in the region around the *x* axis. In contrast, on the SiO$_2$ substrate (Figure S19b), wavefronts are hyperbolic along the *x* direction and elliptic along the *y* direction around the antenna, indicating a closed concave shape of dispersion contour near the *x* axis. These measured results are consistent with our previous experimental results shown in Figures 2 and 3, and also match quite well with the isofrequency contours shown in Figure 4b.

To experimentally demonstrate the concept, we prepare the substrate composed of an Au-SiO$_2$-Au in-plane sandwich structure and transfer the α-MoO$_3$ slab onto the substrate with a deterministic alignment (Figure S20). We excite hybrid polaritons in the heterostructures (with $E_F$=0.5 eV graphene doping) on top of the left gold substrate by an antenna. The near-field distribution is shown in Figure 4c (see more details in Figure S21). When the polaritons of elliptical-wavefronts on top of the left gold substrate cross the boundary between gold and SiO$_2$ substrates, the Poynting vector of the polaritons refracts on the same side of the normal direction, therefore producing what is known as negative refraction, due to the change in the detailed shape of the dispersion contour, which ultimately leads to a partial focusing of the incident polaritons. We observe the formation of a focal spot close to the right Au-SiO$_2$ interface. In Figure S22, numerically simulated in-plane field distributions (Re {Ez}) further corroborate the experimental findings. The red curve in Figure 4d shows the spatial distribution of the electric field amplitude at the focal spot, which demonstrates a significant wavelength compression towards a FWHM of 640 nm along the *y* direction. The simulated spatial distribution of the electrical field (blue curve in Figure 4d) and propagation profile (blue curve in Figure 4e) match well with the experimental results (red curves therein). Note that, by varying the illumination frequency (Figure S23), the anomalous focus spot can be tuned from 660 nm to 500 nm with frequency $\omega$ moving from 893 cm$^{-1}$ to 910 cm$^{-1}$; these distances are less than 1/17 to 1/22 of the corresponding illumination wavelengths, thus emphasizing a deep-subwavelength focusing.

## Conclusion

In conclusion, we have experimentally demonstrated that the topology of the isofrequency dispersion contours for the hybrid polaritons supported in a heterostructure composed of a graphene sheet on top of an α-MoO$_3$ layer can be significantly modified by chemically changing the doping level of graphene, with the contour topology being transformed from open to closed shapes over a broad frequency range. A flat dispersion contour appears at the topological transition, which supports a highly directive and diffractionless polariton propagation, resulting in a tunable canalization mode controlled by the doping level of graphene. Potentially, electrical gating could be used to control the doping level in future studies. Furthermore, by choosing the substrate for the heterostructure, one can also engineer the dispersion contour to have even flattening dispersion contours when a gold substrate is used. This further allows us to design an extremely subwavelength device for in-plane focusing of hybrid polaritons exploiting substrate engineering, where negative refractive occurs at the boundary between two different substrates. Our study provides an avenue towards tunable topological photonics/polaritonic transitions in low-dimensional materials,[31,32] with potential application in optical imaging, sensing, and control of spontaneous emission.

## Methods

**Nanofabrication of the devices.**

High-quality α-MoO$_3$ flakes were mechanically exfoliated from bulk crystals synthesized by the chemical vapor deposition (CVD) method[20] and then transferred onto either commercial 300 nm SiO$_2$/500 μm Si wafers (SVM, Inc.) or gold substrates via a deterministic dry transfer process with a PDMS stamp. CVD-grown monolayer graphene on copper foil was transferred onto α-MoO$_3$ samples using the PMMA-assisted method following our previous report[33]. Gold antenna (3 μm × 50 nm) arrays were patterned on the selected α-MoO$_3$ flakes using 100 kV electron-beam lithography (EBL) (Vistec 5000+ES, Germany) on approximately 350 nm of PMMA950K electron beam lithography resist. Electron-beam evaporation was subsequently used to deposit 5 nm Ti and 50 nm Au in a vacuum chamber at a pressure of <5×10$^{-6}$ Torr to fabricate Au antennas. Electron-beam evaporation was also used to deposit a 60-nm-thick gold film onto a low-doped Si substrate. To remove any residual organic materials, samples were immersed in a hot acetone bath at 80 °C for 25 min and subject to a gentle rinse of IPA for 3 min, followed by nitrogen gas drying and thermal baking. The samples are annealed in a reductive atmosphere to remove most of the dopants from the wet transfer process and are then put in a chamber filled with NO$_2$ gas to achieve different doping levels by surface adsorption of gas molecules. The graphene Fermi energy can then be controlled by playing with the gas concentration and doping time, so it can be shifted up to values as high as ~0.9 eV (Supplementary Materials Figure S7).

**Scanning near-field optical microscopy (SNOM) measurements.**

A scattering SNOM setup (Neaspec GmbH) equipped with wavelength-tunable lasers (between 890 and 2000 cm$^{-1}$) was employed to image optical near-fields. The AFM tip of the microscope was coated by gold with an apex radius of ~25 nm (NanoWorld), and the tip-tapping frequency and amplitudes were set to ~270 kHz and ~30-50 nm, respectively. The p-polarized mid-infrared beam from a tunable $CO_2$ laser was directed to the AFM tip, with lateral spot sizes of ~15 μm under the tip to cover the antennas and a large area of the graphene/α-$MoO_3$ samples. Third-order harmonic demodulation was applied to the near-field amplitude images to strongly suppress background noise.

**Calculation of polariton dispersion and IFCs of hybrid plasmon-phonon polaritons.**

The transfer matrix formalism is adopted to calculate the dispersion and IFCs of the hybrid plasmon-phonon polaritons in graphene-α-$MoO_3$ heterostructures. Our theoretical model mainly includes a three-layer structure: Layer 1 (z>0, air) is a cover layer; Layer 2 (0>z>-$d_h$, graphene/α-$MoO_3$) is a middle layer; and Layer 3 (z<-$d_h$, $SiO_2$ or Au) is a substrate (see Figure S24). Each layer is regarded as a dielectric layer represented by the dielectric tensor. The air and the substrate layer are modeled by isotropic tensors diag{$\epsilon_{a,s}$} [34]. The α-$MoO_3$ slab is modeled by an anisotropic diagonal tensor diag{$\epsilon_{xx}$, $\epsilon_{yy}$, $\epsilon_{zz}$}, where $\epsilon_{xx}$, $\epsilon_{yy}$, and $\epsilon_{zz}$ are the permittivity components along the *x*, *y*, and *z* axes, respectively*(25)*. Also, monolayer graphene is located on the top of α-$MoO_3$ at *z*=0 and described as a zero-thickness current layer characterized by a frequency-dependent surface conductivity taken from the local random-phase approximation model[35]:

$$\sigma(\omega) = \frac{ie^2 K_B T(\omega+1/\tau)}{\pi \hbar^2}\left[\frac{E_F}{K_B T} + 2\ln\left(\exp(-\frac{E_F}{K_B T})+1\right)\right] + i\frac{e^2}{4\pi\hbar}\ln\left[\frac{2|E_F|-\hbar(\omega+i/\tau)}{2|E_F|+\hbar(\omega+i/\tau)}\right], \quad (S1)$$

which depends on the Fermi energy $E_F$, the inelastic relaxation time $\tau$, the temperature *T*, and the relaxation time $\tau = \mu E_F / e v_F^2$, expressed in turn in terms of the graphene Fermi velocity $v_F = c/300$ and the carrier mobility $\mu$.

Given the strong confinement of the structure under consideration, we only need to consider TM modes, since TE components contribute negligibly. The corresponding p-polarization Fresnel reflection coefficient $r_p$ of the three-layer system under consideration admits the analytical expression

$$r_p = \frac{r_{12} + r_{23}(1 - r_{12} - r_{21})e^{i2k_z^{(2)}d_h}}{1 + r_{12}r_{23}e^{i2k_z^{(2)}d_h}} \quad (S2)$$

$$r_{12} = \frac{Q_1 - Q_2 + SQ_1Q_2}{Q_1 + Q_2 + SQ_1Q_2} \quad (S3)$$

$$r_{21} = \frac{Q_2 - Q_1 + SQ_1Q_2}{Q_2 + Q_1 + SQ_1Q_2} \tag{S4}$$

$$r_{23} = \frac{Q_2 - Q_3}{Q_2 + Q_3} \tag{S5}$$

$$Q_j = \frac{k_z^{(j)}}{\epsilon_t^{(j)}} \tag{S6}$$

$$S = \frac{\sigma Z_0}{\omega} \tag{S7}$$

Here, $r_{jk}$ denotes the reflection coefficient of the *j-k* interface for illumination from medium *j*, with *j, k*=1-3; $\epsilon_t^{(j)}$ is the in-plane dielectric function for a propagation wave vector $k_p(\theta)$ (with $\theta$ being the angle to the *x* axis), which can be expressed as $\epsilon_t^{(j)} = \epsilon_x^{(j)} \cos^2(\theta) + \epsilon_y^{(j)} \sin^2(\theta)$, where $\epsilon_x^{(j)}$ and $\epsilon_y^{(j)}$ are the dielectric functions of layer j along the *x* and *y* axes, respectively; $k_z^{(j)} = \sqrt{\varepsilon_t^{(j)} \frac{\omega^2}{c^2} - \frac{\varepsilon_t^{(j)}}{\varepsilon_z^{(j)}} q^2}$ is the out-of-plane wave vector, where $\epsilon_z^{(j)}$ is the dielectric function of layer *j* along the *z* axes; and $Z_0$ is the vacuum impedance.

We can find the polariton dispersion relation $q(\omega, \theta)$ from the zeros of the denominator of Eq. (S2),

$$1 + r_{12}r_{23}e^{i2k_z^{(2)}d_h} = 0. \tag{S8}$$

For simplicity, we consider a system with small dissipation, so the maxima of Im$\{r_p\}$ (see color plots in Figure 2M-T) approximately solve the condition given by Eq. (S8), and therefore, the produce the sought-after dispersion relation $q(\omega, \theta)$.

**Electromagnetic simulations.**

The electromagnetic fields around the antennas were calculated using a Finite-Elements Method package (COMSOL). The metallic antennas were placed on a graphene/α-MoO$_3$ sample, and this in turn on 300 nm SiO$_2$ or 60 nm gold substrates, which were illuminated by a light plane wave, incident under 45° relative to the surface and with surface-projected polarization parallel to the long antenna axis. We also used a dipole polarized along the *z* direction to launch polaritons, and the distance between the dipole and the uppermost surface of the sample was set to 100 nm.

We obtained the distribution of Re {$E_Z$} (real part of the out-of-plane electric field) over a plane placed 20 nm above the surface of graphene. The boundary conditions were set to perfectly matched layers. Graphene was modeled as a transition interface with a conductivity described by the local random-phase approximation model. We assumed a graphene carrier mobility of 2000 cm$^2$/(V·s). The permittivity of silicon dioxide and Au at the used mid-infrared wavelength are shown in Figs. S1c and S1d, respectively, in the Supplementary Materials.

## Acknowledgments

The authors acknowledge Dr. Pablo Alonso-González and Dr. Jiahua Duan (Departamento de Física, Universidad de Oviedo) for valuable discussions and constructive comments. This work was supported by the National Key Research and Development Program of China (Grant No. 2020YFB2205701), the National Natural Science Foundation of China (Grant Nos. 51902065, 52172139, 51925203, U2032206, 52072083, and 51972072), Beijing Municipal Natural Science Foundation (Grant No. 2202062), and Strategic Priority Research Program of Chinese Academy of Sciences (Grant No. XDB36000000, XDB30000000). F.J.G.A. acknowledges the ERC (Advanced Grant 789104-eNANO), the Spanish MINECO (SEV2015-0522), and the CAS President's International Fellowship Initiative (PIFI) for 2021. S. F. acknowledges the support of the U.S. Department of Energy under Grant No. DE-FG02-07ER46426.

## Data availability

The data that support the findings of this study are available from the corresponding author upon reasonable request.

## Author contributions

Q.D., R.Y., H.H., and F.J.G.A. conceived the idea. Q.D., F.J.G.A., and S.F. supervised the project. H.H. and N.C. led the experiments. R.Y., H.T., and F.J.G.A. developed the theory and performed the simulation. H.H. and N.C. prepared the samples and performed the near-field measurements. H.H., R.Y., N.C., and H.T. analyzed the data, and all authors discussed the results. R.Y. and H.H. wrote the manuscript, with input and comments from all authors.

## Competing interests

The authors declare no competing interests.

**Note added in proof:** During preparing the manuscript, a related theoretical study of tuning of highly anisotropic phonon polaritons in graphene and α-MoO3 van der Waals structure was reported.[36]

# Reference

1. Kappera, R. *et al.* Phase-engineered low-resistance contacts for ultrathin $MoS_2$ transistors. *Nat. Mater.* **13**, 1128-1134 (2014).
2. Li, Y., Duerloo, K.-A. N., Wauson, K. & Reed, E. J. J. N. c. Structural semiconductor-to-semimetal phase transition in two-dimensional materials induced by electrostatic gating. *Nat. Commun.* **7**, 1-8 (2016).
3. Wang, Y. *et al.* Structural phase transition in monolayer $MoTe_2$ driven by electrostatic doping. *Nature* **550**, 487-491 (2017).
4. Zheng, Y.-R. *et al.* Doping-induced structural phase transition in cobalt diselenide enables enhanced hydrogen evolution catalysis. *Nat. Commun.* **9**, 2533-2539 (2018).
5. Zhang, F. *et al.* Electric-field induced structural transition in vertical $MoTe_2$- and $Mo_{1-x}W_xTe_2$-based resistive memories. *Nat. Mater.* **18**, 55-61 (2019).
6. Huang, M. *et al.* Voltage control of ferrimagnetic order and voltage-assisted writing of ferrimagnetic spin textures. *Nat. Nanotech.* **16**, 1-8 (2021).
7. Walter, J. *et al.* Voltage-induced ferromagnetism in a diamagnet. *Sci. Adv.* **6**, eabb7721 (2020).
8. Zheng, L. M. *et al.* Ambipolar ferromagnetism by electrostatic doping of a manganite. *Nat. Commun.* **9**, 1897-1897 (2018).
9. Jiang, S., Li, L., Wang, Z., Mak, K. F. & Shan, J. Controlling magnetism in 2D $CrI_3$ by electrostatic doping. *Nat. Nanotech.* **13**, 549-553 (2018).
10. Walter, J. *et al.* Giant electrostatic modification of magnetism via electrolyte-gate-induced cluster percolation in $La_{1-x}Sr_xCoO_{3-\delta}$. *Phys. Rev. Mater.* **2**, 111406 (2018).
11. Kim, D. *et al.* Tricritical Point and the Doping Dependence of the Order of the Ferromagnetic Phase Transition of $La_{1-x}Ca_xMnO_3$. *Phys. Rev. Lett.* **89**, 227202 (2002).
12. Chen, B. *et al.* Intrinsic magnetic topological insulator phases in the Sb doped $MnBi_2Te_4$ bulks and thin flakes. *Nat. Commun.* **10**, 1-8, (2019).
13. Sajadi, E. *et al.* Gate-induced superconductivity in a monolayer topological insulator. *Science* **362**, 922-925 (2018).
14. Chen, Z. *et al.* Carrier density and disorder tuned superconductor-metal transition in a two-dimensional electron system. *Nat. Commun.* **9**, 1-6 (2018).
15. Liu, Q., Zhang, X., Abdalla, L. B., Fazzio, A. & Zunger, A. Switching a normal insulator into a topological insulator via electric field with application to phosphorene. *Nano Lett.* **15**, 1222-1228 (2015).
16. Chen, Y. *et al.* Phase engineering of nanomaterials. *Nat. Rev. Chem.* **4**, 243-256 (2020).
17. Choi, Y. *et al.* Correlation-driven topological phases in magic-angle twisted bilayer graphene. *Nature* **589**, 536-541 (2021).
18. Gomez-Diaz, J. S., Tymchenko, M. & Alù, A. Hyperbolic Plasmons and Topological Transitions Over Uniaxial Metasurfaces. *Phys. Rev. Lett.* **114**, 233901-233901 (2015).
19. Yu, R., Alaee, R., Boyd, R. W. & de Abajo, F. J. G. Ultrafast Topological Engineering in Metamaterials. *Phys. Rev. Lett.* **125**, 037403 (2020).
20. Chen, M. et al. Configurable phonon polaritons in twisted α-$MoO_3$. *Nat. Mater.* **19**, 1307-1311 (2020).
21. Duan, J. *et al.* Twisted Nano-Optics: Manipulating Light at the Nanoscale with Twisted Phonon Polaritonic Slabs. *Nano Lett.* **20**, 5323-5329 (2020).

22    Hu, G. *et al.* Topological polaritons and photonic magic angles in twisted α-MoO$_3$ bilayers. *Nature* **582**, 209-213 (2020).
23    Zheng, Z. *et al.* Phonon Polaritons in Twisted Double-Layers of Hyperbolic van der Waals Crystals. *Nano Lett.* **20**, 5301-5308 (2020).
24    Zhang, Q. *et al.* Hybridized hyperbolic surface phonon polaritons at α-MoO$_3$ and polar dielectric interfaces. *Nano Lett.* **21**, 3112-3119 (2021).
25    Duan, J. *et al.* Enabling propagation of anisotropic polaritons along forbidden directions via a topological transition. *Sci. Adv.* **7**, eabf2690 (2021).
26    Ma, W. *et al.* In-plane anisotropic and ultra-low-loss polaritons in a natural van der Waals crystal. *Nature* **562**, 557-562 (2018).
27    Zheng, Z. *et al.* A mid-infrared biaxial hyperbolic van der Waals crystal. *Sci. Adv.* **5**, eaav8690, (2019).
28    Vakil, A. & Engheta, N. J. S. Transformation optics using graphene. *Science* **332**, 1291-1294 (2011).
29    Davoyan, A. & Engheta, N. J. N. c. Electrically controlled one-way photon flow in plasmonic nanostructures. *Nat. Commun.* **5**, 1-5 (2014).
30    Basov, D., Fogler, M. & De Abajo, F. G. J. S. Polaritons in van der Waals materials. *Science* **354**, 6309 (2016).
31    West, P. R. *et al.* Searching for better plasmonic materials. *Laser Photonics Rev.* **4**, 795-808 (2010).
32    Naik, G. V., Shalaev, V. M. & Boltasseva, A. J. A. M. Alternative plasmonic materials: beyond gold and silver. *Adv. Mater.* **25**, 3264-3294 (2013).
33    Hu, H. *et al.* Far-field nanoscale infrared spectroscopy of vibrational fingerprints of molecules with graphene plasmons. *Nat. Commun.* **7**, 12334-12334 (2016).
34    Kischkat, J. *et al.* Mid-infrared optical properties of thin films of aluminum oxide, titanium dioxide, silicon dioxide, aluminum nitride, and silicon nitride. *Appl. Opt.* **51**, 6789-6798 (2012).
35    Wunsch, B., Stauber, T., Sols, F. & Guinea, F. J. N. J. o. P. Dynamical polarization of graphene at finite doping. *New J. Phys.* **8**, 318 (2006).
36    Álvarez-Pérez, G. *et al.* Active tuning of highly anisotropic phonon polaritons in van der Waals crystal slabs by gated graphene. *preprint arXiv*:2110.11683 (2021).

# Supplementary Materials

**This PDF file includes:**
**Figure S1.** Optical parameters of the materials in this work.
**Figure S2.** Theoretically calculated isofrequency contours of hybrid polaritons at different illumination frequencies and Fermi energies of graphene.
**Figure S3.** Numerically simulated field distribution of hybrid polaritons on a 300 nm $SiO_2$ substrate with different Fermi energies of graphene.
**Figure S4.** Numerically simulated field distribution of hybrid polaritons on a 60 nm gold substrate with different Fermi energies of graphene.
**Figure S5.** Optical image of graphene/α-MoO3 samples.
**Figure S6.** Schematics of s-SNOM nano-imaging of hybrid polaritons.
**Figure S7.** Real-space infrared nano images reveal hybrid polaritons on the 300 nm $SiO_2$ substrate with different Fermi energies of graphene and various illumination frequencies.
**Figure S8.** Real-space infrared nano images reveal hybrid polaritons on the 60 nm Au substrate with different Fermi energies of graphene and various illumination frequencies.
**Figure S9.** Chemical doping and Raman spectra of the graphene/α-MoO3 heterostructure.
**Figure S10.** Dispersions of hybrid polaritons.
**Figure S11.** Polariton canalizations near the topological transition at different illumination frequencies.
**Figure S12.** The full width at half maximum (FWHM) of polariton canalizations.
**Figure S13.** Real-space infrared nano images reveal hybrid polaritons in a sample with disorder.
**Figure S14.** Numerically simulated field distribution of the $z$ component of hybrid polaritons.
**Figure S15.** Optical image of a graphene/α-MoO3 heterostructure on a gold substrate with different angles of the antenna.
**Figure S16.** Extraction analysis of antenna-tailored launching of hybrid polaritons.
**Figure S17.** Numerically simulated field distribution of hybrid polaritons launched by gold antennas with different antenna angles.
**Figure S18.** Numerically simulated negative refraction of hybrid polaritons with different Fermi energies of graphene.
**Figure S19.** Experimentally measured polariton near-field distributions for samples on 60-nm-thick gold and 300-nm-thick $SiO_2$ substrates.
**Figure S20.** Optical and topography images of a negative refraction lens sample.
**Figure S21.** Method to extract antenna-launched hybrid polaritons.
**Figure S22.** Numerically simulated field distributions (Re {Ez}) compared with the measured results shown in Figure 4c.
**Figure S23.** Partial focusing of hybrid polaritons by a negative refraction lens for different illumination frequencies.
**Figure S24.** Illustration of the geometry considered for the theoretical model.
**References**

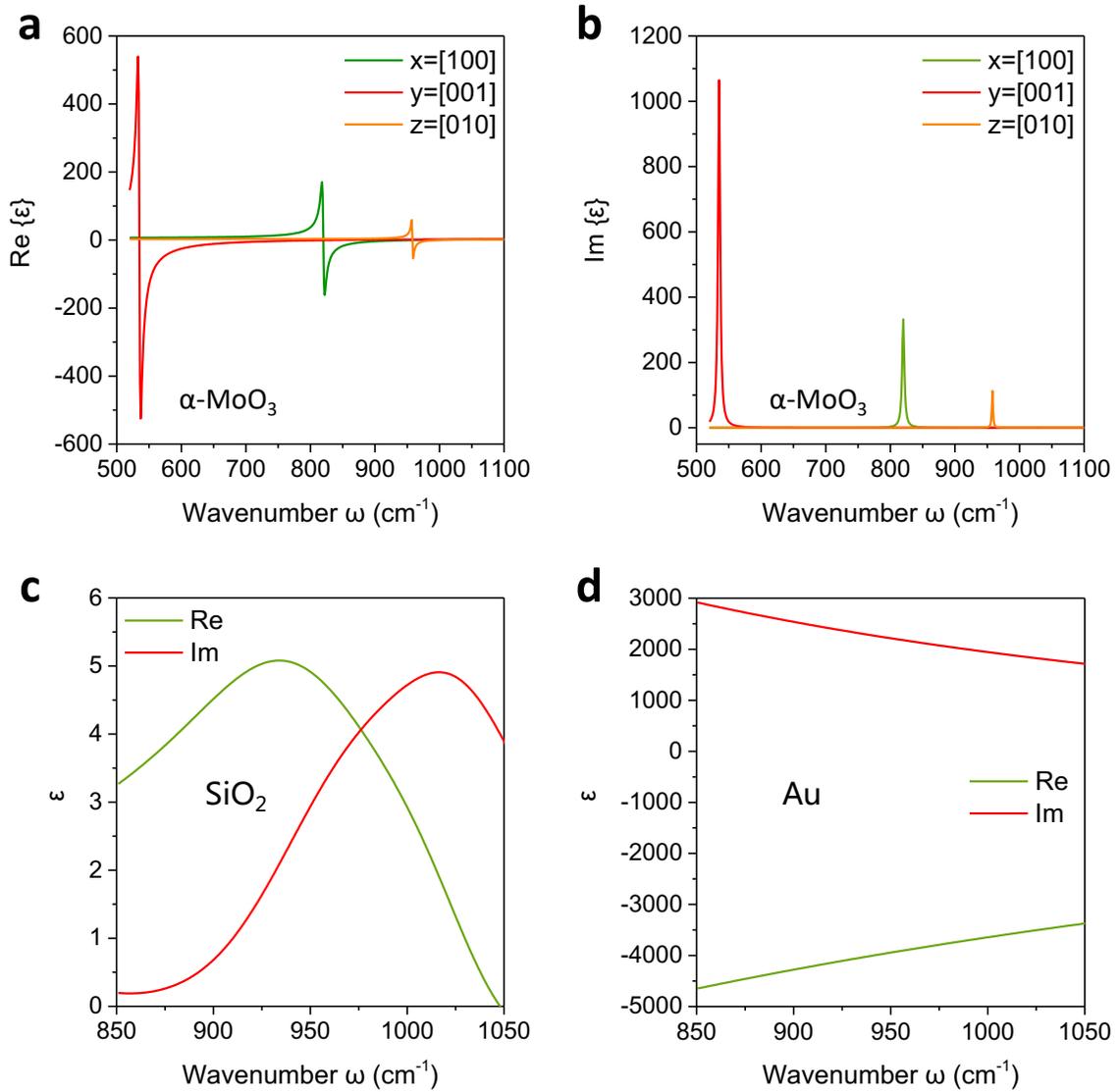

**Figure S1. Optical parameters of the materials in this work. (a, b)** Permittivity of α-MoO$_3$ along different principal directions. The permittivity is obtained by fitting the obtained data with the Lorentzian model.[1] **(c, d)** Permittivity of SiO$_2$ and gold.[2, 3]

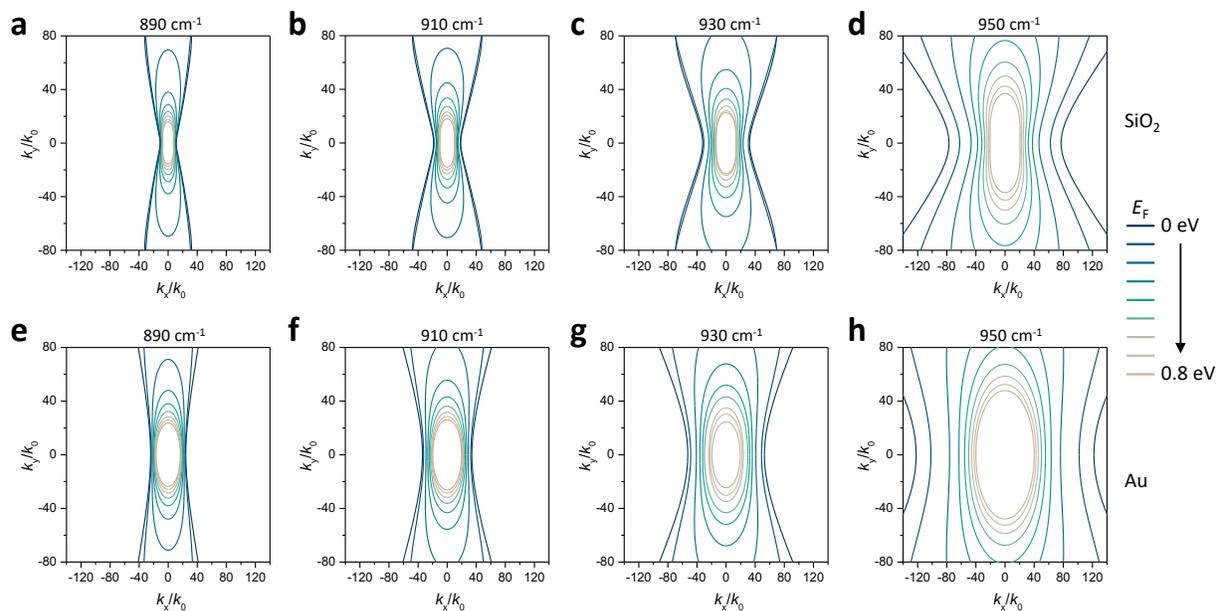

**Figure S2. Theoretically calculated isofrequency contours of hybrid polaritons at different illumination frequencies and Fermi energies of graphene. (a-d)** 300 nm $SiO_2$ substrate. **(e-h)** 60 nm gold substrate.

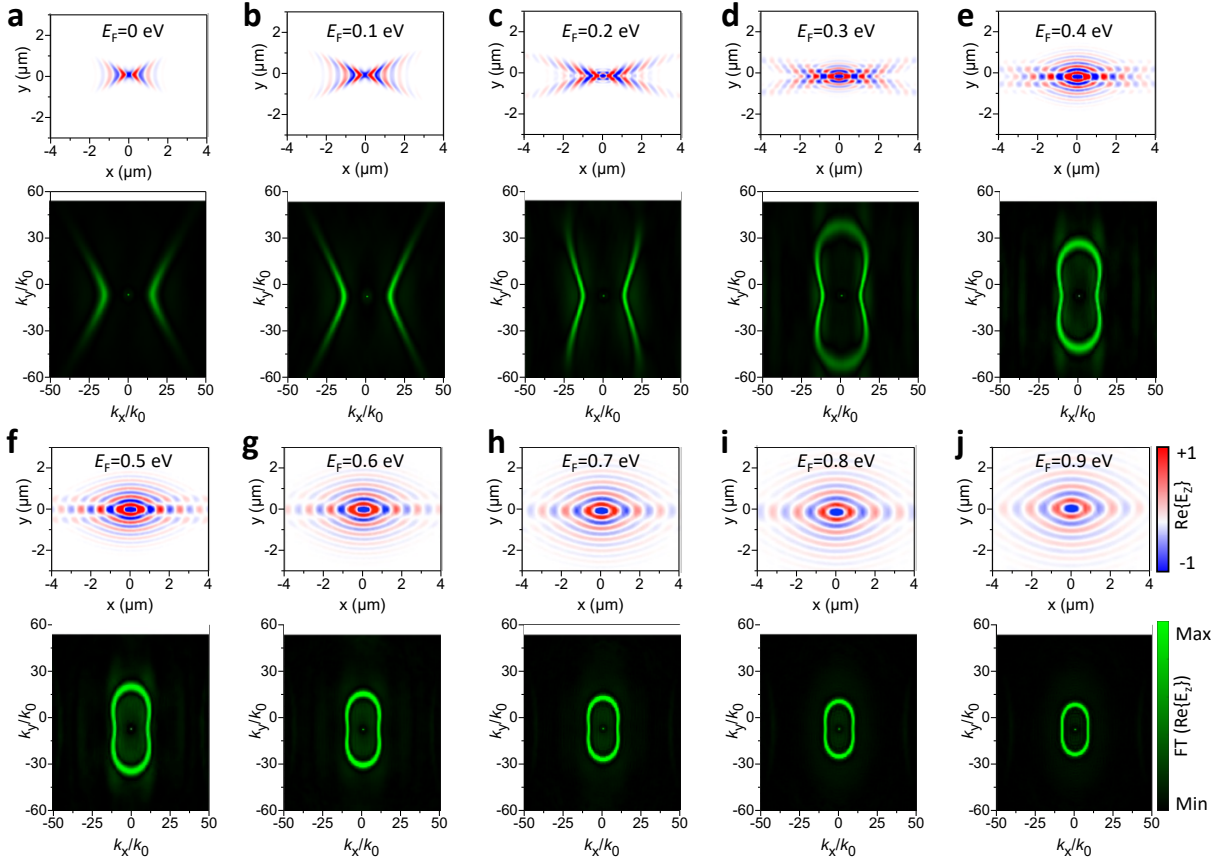

**Figure S3. Numerically simulated field distribution of hybrid polaritons on a 300 nm SiO$_2$ substrate with different Fermi energies of graphene.** A dipole is placed 100 nm above the graphene to launch the polaritons. The electric distribution is obtained over a plane situated 20 nm above the graphene. The graphene Fermi energy is varied from 0 to 0.9 eV (see labels), the thickness of α-MoO$_3$ is 150 nm, and the incident light wavelength is fixed at $\lambda_0$ = 10.99 μm (910 cm$^{-1}$) in all simulations.

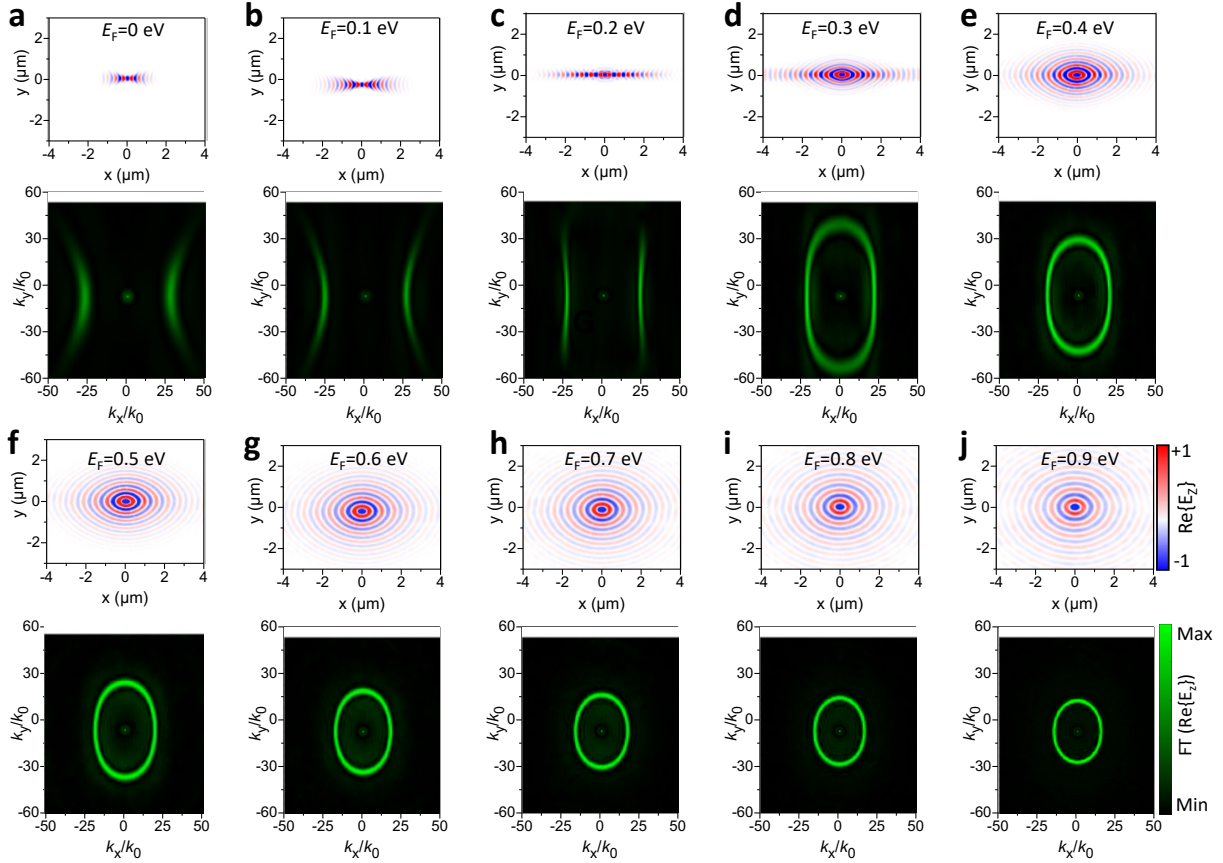

**Figure S4. Numerically simulated field distribution of hybrid polaritons on a 60 nm gold substrate with different Fermi energies of graphene.** A dipole is placed 100 nm above the graphene to launch the polaritons. The electric distribution is obtained over a plane situated 20 nm above the graphene. The graphene Fermi energy is varied from 0 to 0.9 eV (see labels), the thickness of α-MoO$_3$ is 150 nm, and the incident light wavelength is fixed at $\lambda_0$ = 10.99 μm (910 cm$^{-1}$) in all simulations.

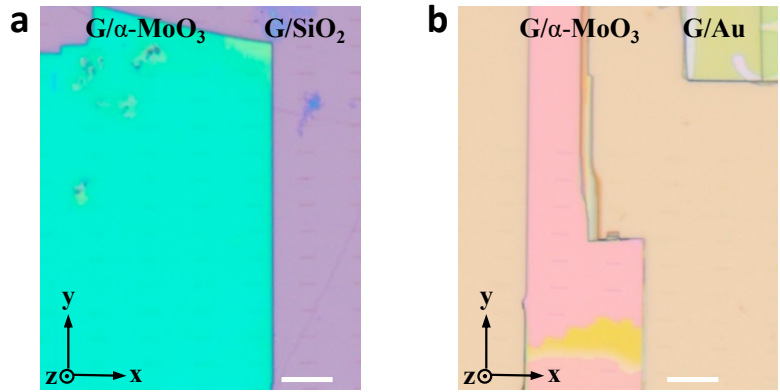

**Figure S5. Optical image of graphene/α-MoO$_3$ heterostructures.** The sample is supported by the 300 nm SiO$_2$ substrate (**a**, α-MoO$_3$ thickness d=140 nm) and 60 nm gold substrate (**b**, α-MoO$_3$ thickness d=140 nm) with a gold antenna array. The scale bar indicates 10 μm.

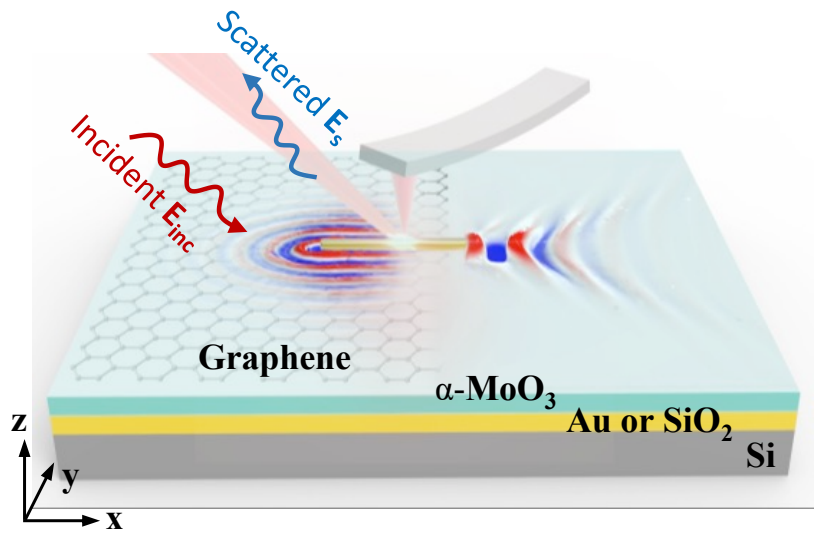

**Figure S6. Schematics of s-SNOM nano-imaging of polaritons.** A gold antenna (yellow) confines infrared light that allows launching polaritons in the sample, which are probed by a metalized AFM tip and then scattered into free space for collection by a distant detector.

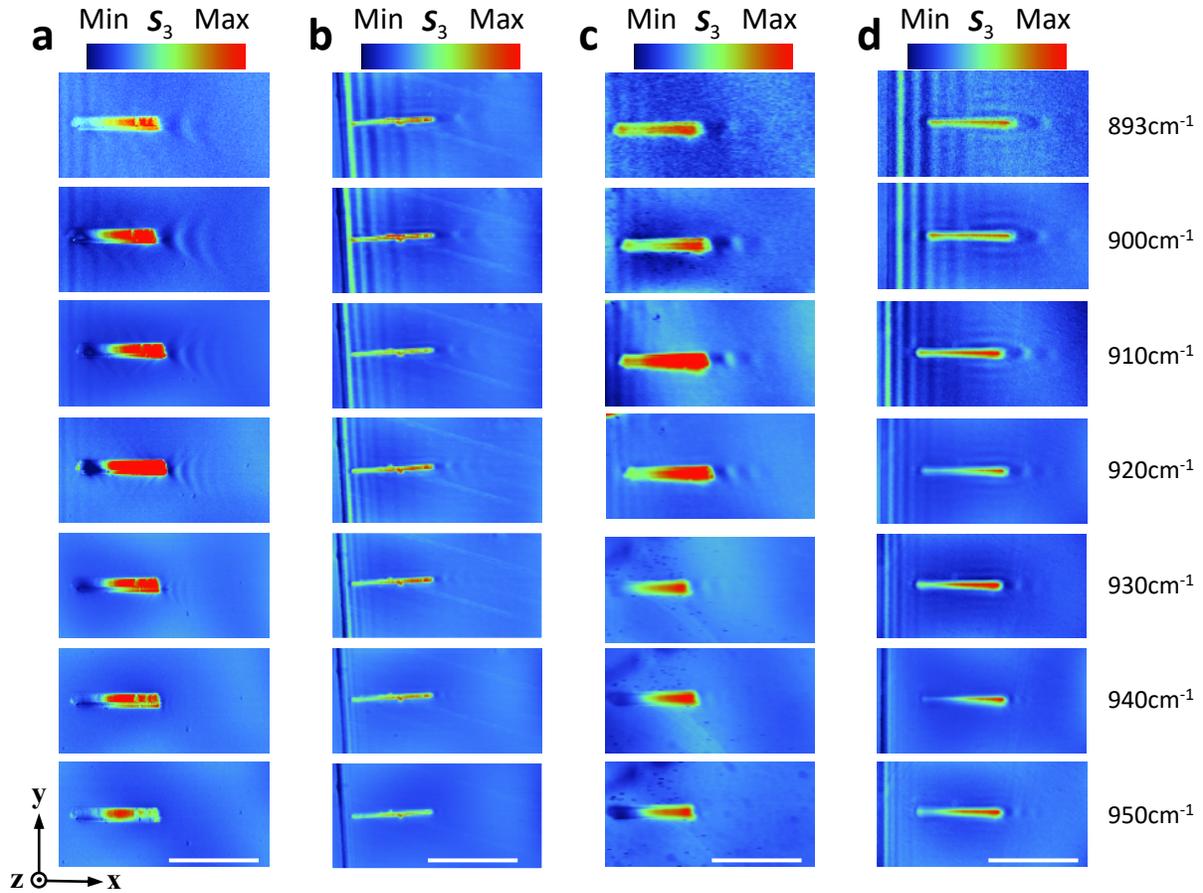

**Figure S7. Real-space infrared nano images reveal hybrid polaritons on the 300 nm SiO$_2$ substrate with different Fermi energies of graphene and various illumination frequencies.** (**a-d**) Antenna-launched hybrid polaritons measured in a heterostructure sample consisting of a 140-nm-thick α-MoO$_3$ film and monolayer graphene. The Fermi energy of graphene is set at $E_F$=0 eV (**a**), $E_F$=0.3 eV (**b**), $E_F$=0.4 eV (**c**), and $E_F$=0.7 eV (**d**), respectively. The incident light wavelength is tuned from $\lambda_0$ = 11.20 μm (893 cm$^{-1}$) to 10.53 μm (950 cm$^{-1}$) in the experiments. The scale bars indicate 3 μm.

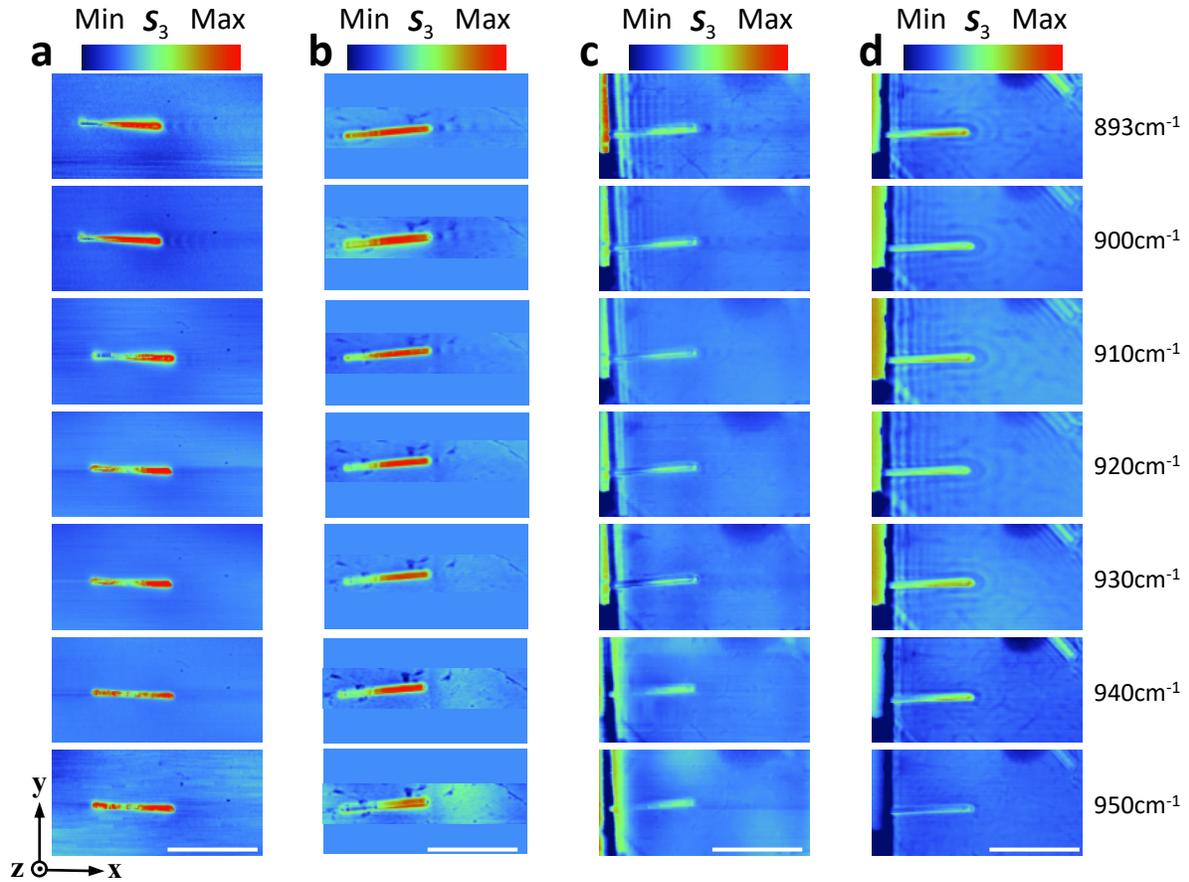

**Figure S8. Real-space infrared nano images reveal hybrid polaritons on the 60 nm Au substrate with different Fermi energies of graphene and various illumination frequencies.** (**a-d**) Antenna-launched hybrid polaritons measured in a heterostructure sample consisting of a 140-nm-thick α-MoO$_3$ slab and monolayer graphene. The Fermi energy of graphene is set at $E_F$=0 eV (**a**), $E_F$=0.3 eV (**b**), $E_F$=0.4 eV (**c**), and $E_F$=0.7 eV (**d**), respectively. The incident light wavelength is tuned from $\lambda_0$ = 11.20 μm (893 cm$^{-1}$) to 10.53 μm (950 cm$^{-1}$) in the experiments. The scale bars indicate 3 μm.

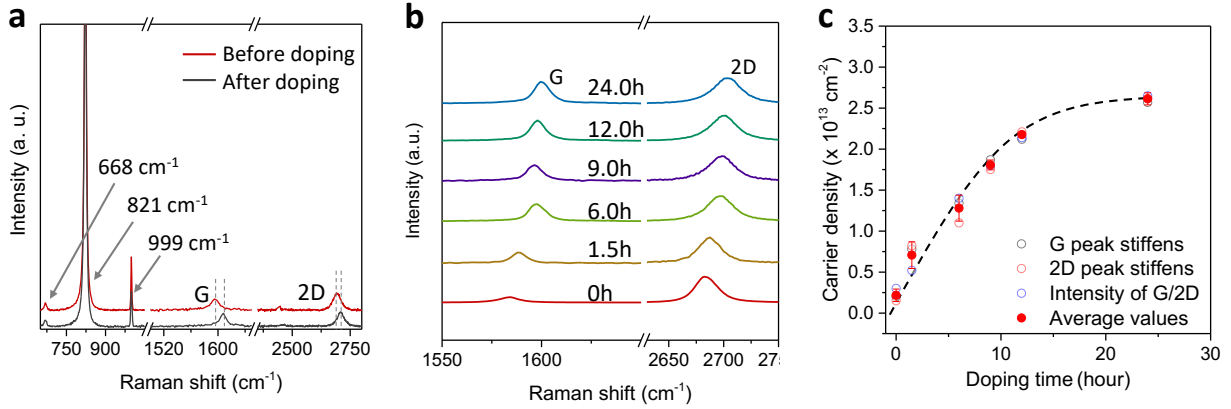

**Figure S9. Chemical doping and Raman spectra of the Graphene/α-MoO₃ heterostructure.**
(**a**) Raman spectra of the G/α-MoO₃ heterostructure obtained before and after doping of graphene. (**b**) Raman spectra of graphene after different doping times. The concentration of NO₂ gas is 75% in a N₂ atmosphere. (**c**) Carrier densities of graphene as a function of doping time. The carrier density is calculated by following three different methods, from the G peak stiffening (grey circles), the 2D peak stiffening (red circles), or the intensity ratio $I_G/I_{2D}$ (blue circles). The solid red circles represent the average values of the above three calculated results. Symbols are obtained from experimental data in (b), whereas the dashed curve is a guide to the eye.

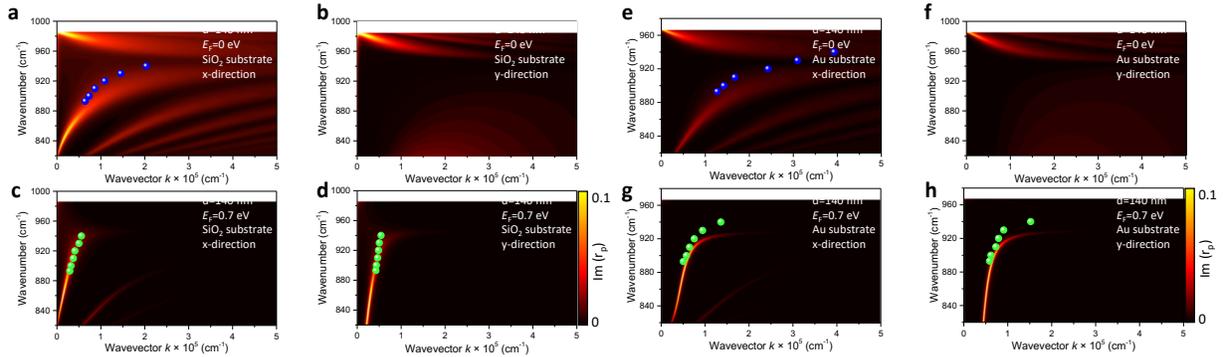

**Figure S10. Dispersions of hybrid polaritons.** Theoretically calculated (color plots) dispersions of hybrid polaritons on the SiO₂ (a-d) or Au (e-h) substrates for wave vector along *x* ([100] direction of α-MoO₃, a, c, e, and g) and *y* ([001] direction of α-MoO₃, b, d, f, and h) with $E_F$=0 eV (a, b, e, and f) and $E_F$=0.7 eV (c, d, g, and h), respectively. Colored symbols are experimental data points extracted from Figures S7 and S8 in the Supplementary Materials.

When graphene is undoped (Figures S10a, b), we find dispersion diagrams similar to those of PhPs supported in α-MoO₃ on SiO₂ substrate, where an intrinsic forbidden band (840-950 cm⁻¹) for polaritons propagating along the y direction can be seen in Figure S10b, which leads to a hyperbolic isofrequency dispersion contour within the band. When graphene is highly doped ($E_F$=0.7 eV, Figures S10c, d), the in-plane hybrid polariton is dominated by a graphene-plasmon character, where a bright feature with a steeper slope appears in the dispersion diagram in both propagating directions. As for the samples on the gold substrate (Figures S10e-h), the hybrid polaritons become more subwavelength due to screening.

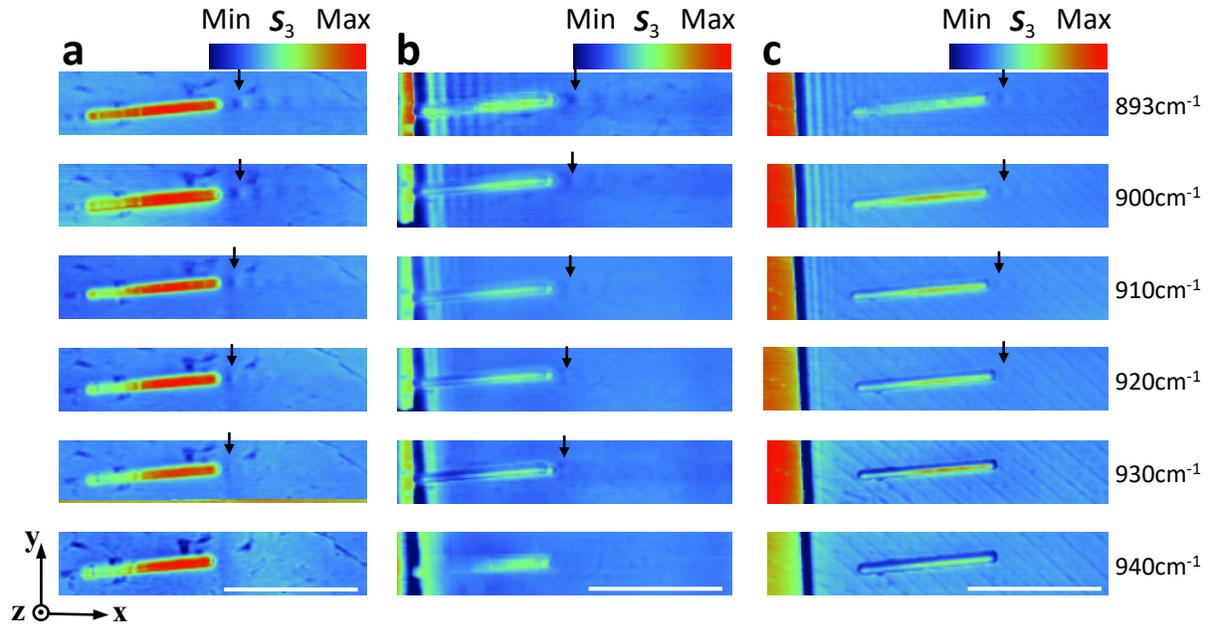

**Figure S11 Polariton canalizations near the topological transition at different illumination frequencies.** (**a**) Images obtained from a sample with the graphene Fermi energy set at $E_F$=0.3 eV, an α-MoO$_3$ thickness of 140 nm, a substrate consisting of a 60-nm-thick gold substrate. (**b**) Same as (a), but for a graphene Fermi energy $E_F$=0.4 eV. (**c**) Same as (b), but for an α-MoO$_3$ thickness of 120 nm. The scale bars indicate 3 μm.

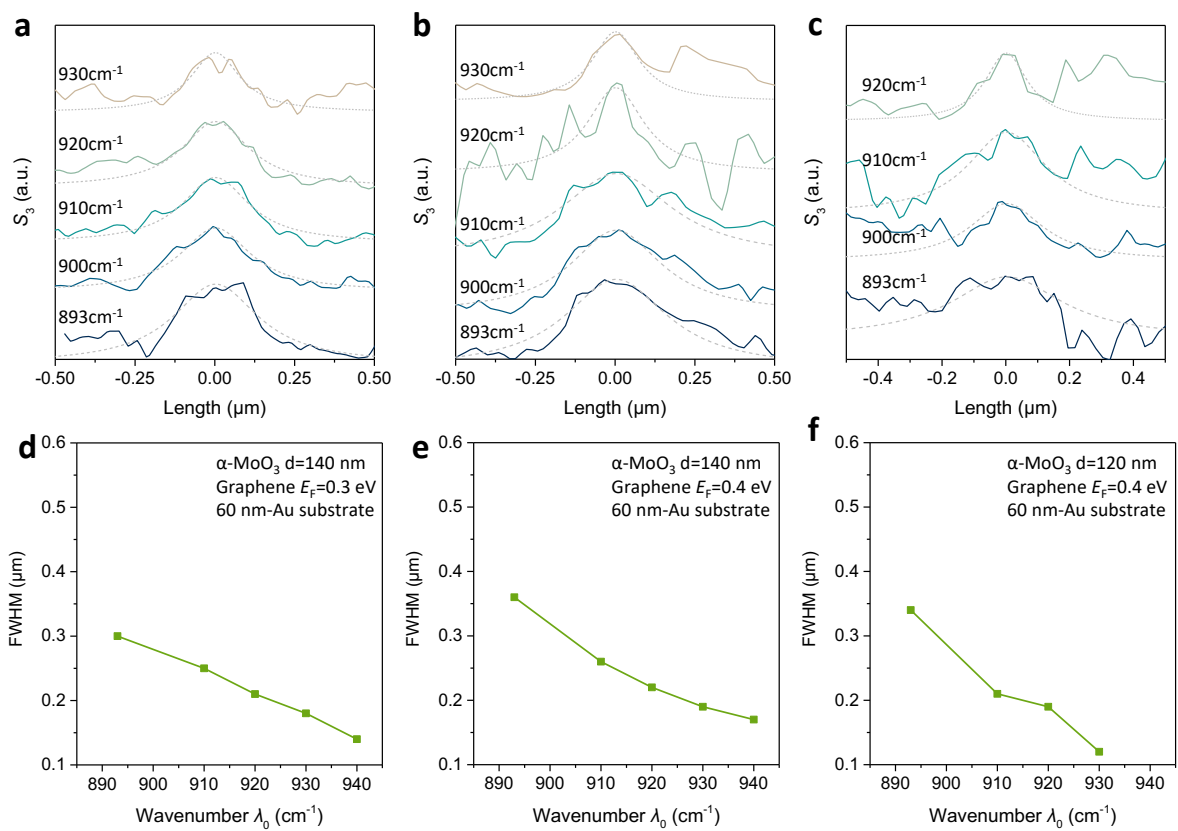

**Figure S12. The full width at half maximum (FWHM) of polariton canalizations.** (**a-c**) Experimental near-field intensity (solid curves) at the black arrows in Figures S11a-c, respectively. The dashed curves show Lorentz fittings of the experimental data. (**d-f**) Dependence of the FWHM of the spot size for polariton canalization as a function of incident light wavelength $\lambda_0$ for samples with different α-MoO$_3$ thickness and substrates (see labels). Green curves serve as a guide to the eye.

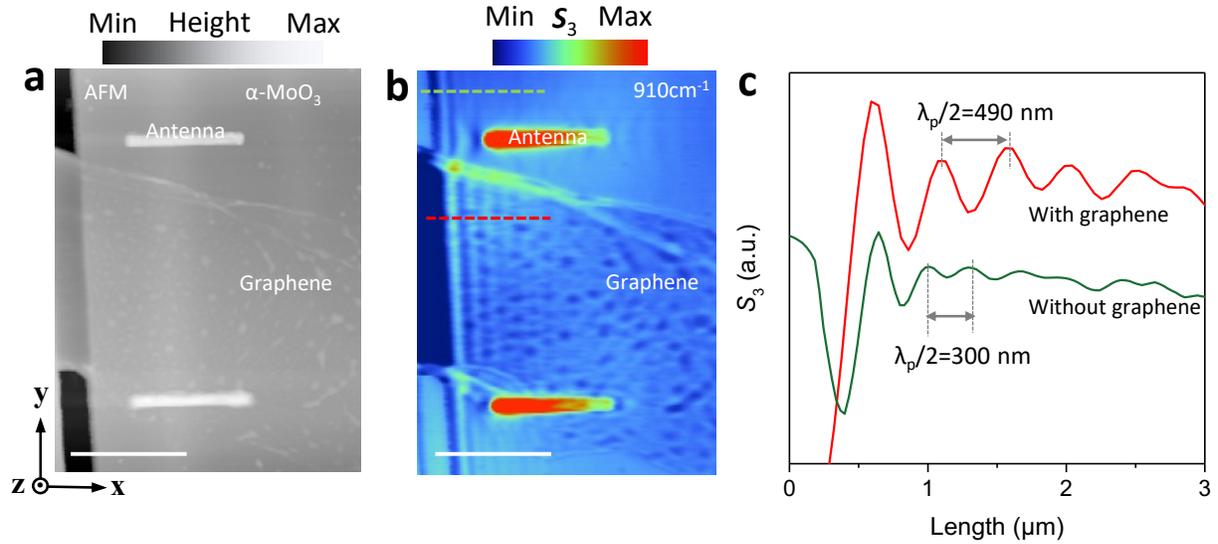

**Figure S13. Real-space infrared nano images reveal hybrid polaritons in a sample with disorder.** (**a**) Topography image showing a graphene monolayer covering half of the α-MoO$_3$ region. (**b**) Near-field image recorded simultaneously with the topography shown in (a). The Fermi energy of graphene is set at $E_F$=0.7 eV, and the α-MoO$_3$ thickness is 300 nm. The incident light wavelength is $\lambda_0$ = 10.99 μm (910 cm$^{-1}$). The scale bar indicates 3 μm. (**c**) Near-field profiles taken at the positions marked by green and red dashed lines in (b). The near-field data are shown in arbitrary units.

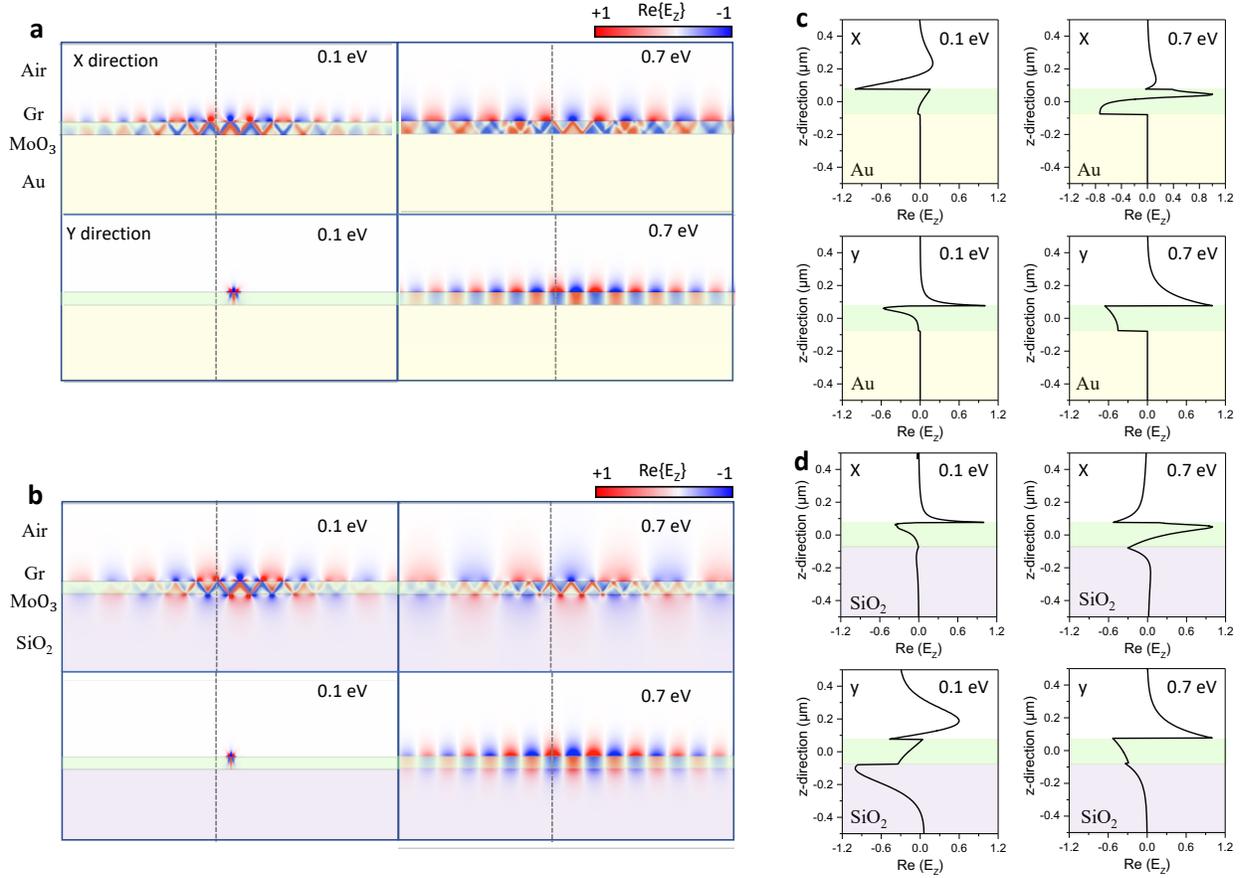

**Figure S14. Numerically simulated field distribution of the *z* component of hybrid polaritons.** (**a, b**) The sample is lying on a 60 nm gold substrate ((Air/G/150 nm α-MoO$_3$/60nm Au), (a)) and 300 nm SiO$_2$ substrate ((Air/G/150 nm α-MoO$_3$/300nm SiO$_2$), (b)). A dipole placed 100 nm above the graphene to launch the polaritons. (**c, d**) Near-field profiles of the real part of the *z* component of the electric field, Re {$E_z$}, perpendicular to the graphene surface. The profiles are extracted along the dashed grey line in (a) and (b). The incident light wavelength is fixed at $\lambda_0$ = 10.99 μm (910 cm$^{-1}$) in all simulations.

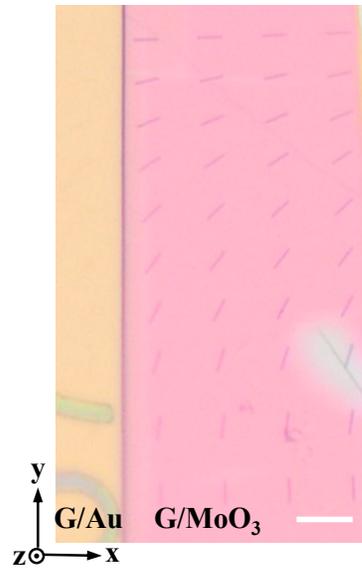

**Figure S15. Optical image of a graphene/α-MoO₃ heterostructure on a gold substrate with different angles of the antenna.** The α-MoO$_3$ thickness is 207 nm. The scale bar indicates 6 μm.

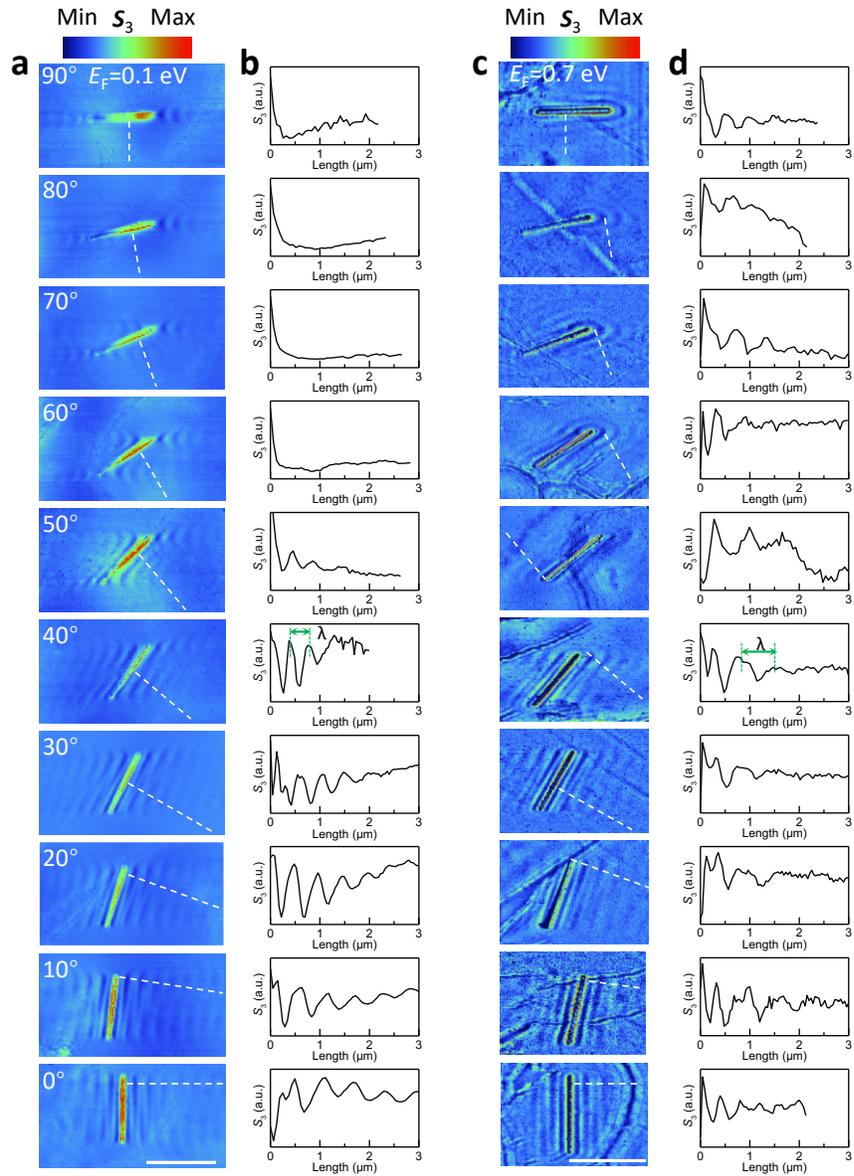

**Figure S16. Extraction analysis of antenna-tailored launching of hybrid polaritons.** The scale bar indicates 3 μm.

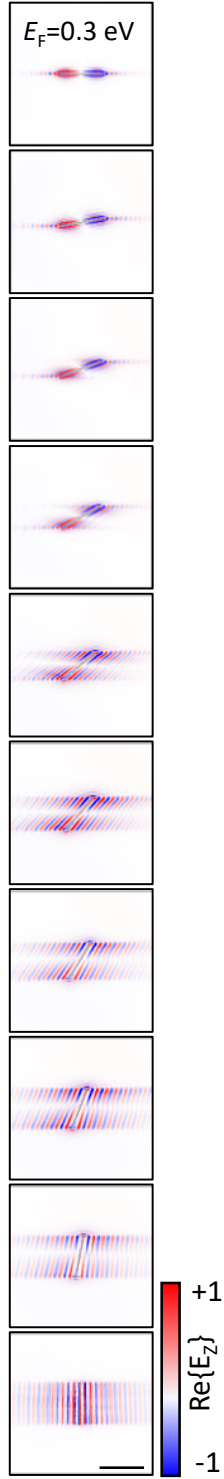

**Figure S17. Numerically simulated field distribution of hybrid polaritons launched by gold antennas with different antenna angles.** The graphene Fermi energy is set at $E_F$=0.3 eV. The thickness of α-MoO$_3$ is 207 nm. The scale bar indicates 3 μm.

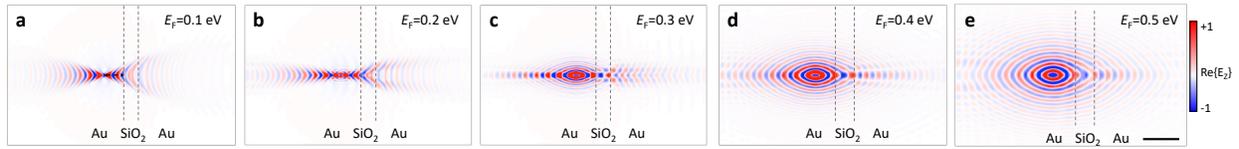

**Figure S18. Numerically simulated negative refraction of hybrid polaritons with different Fermi energies of graphene from $E_F$=0.1 to 0.5 eV.** (a-e) The hybrid polaritons are excited on the 60 nm left Au substrate and propagate first directionally towards the SiO$_2$ area and later to the right Au area. The polaritons are negatively refracted at the boundary and converge to form a focus in the right area when the topological polaritons are launched for high doping of graphene. The dashed grey line represents the interface between Au and SiO$_2$ areas. The scale bar indicates 3 μm.

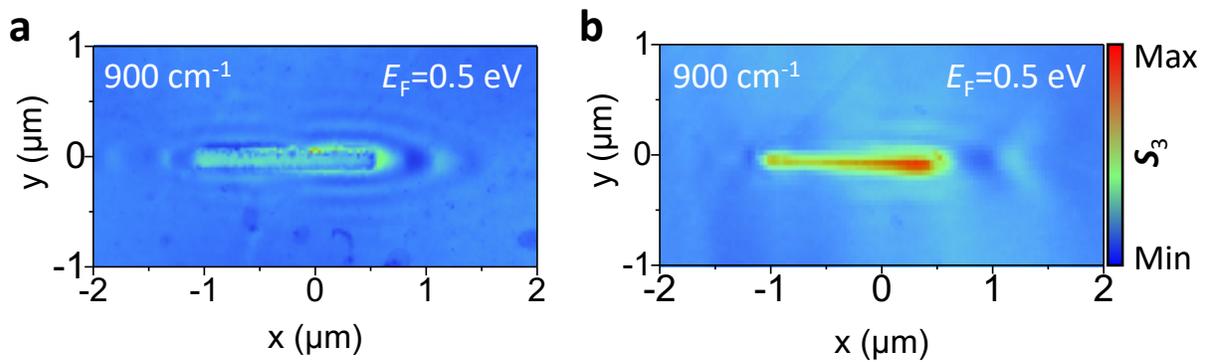

**Figure S19. Experimentally measured polariton near-field distributions for samples on 60-nm-thick gold (a) and 300-nm-thick SiO$_2$ (b) substrates, respectively.** The polaritons are launched by a gold antenna (length 3.2 μm; width 250 nm; thickness 50 nm).

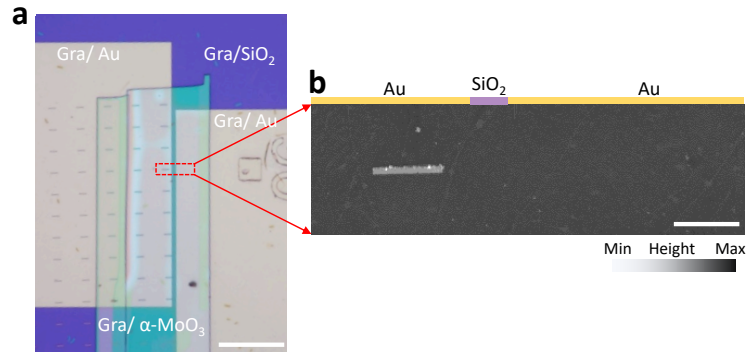

**Figure S20. Optical and topography images of a negative refraction lens sample.** (**a**) Optical image of a negative refraction lens. The scale bar indicates 30 μm. (**b**) Topography image of a negative refraction lens. The scale bar indicates 3 μm.

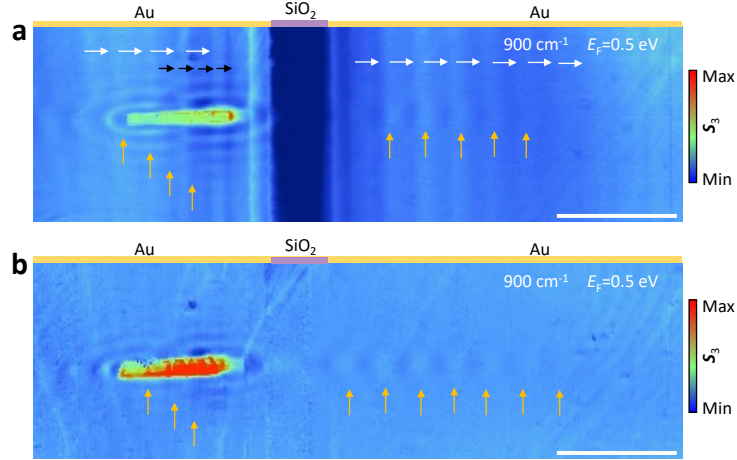

**Figure S21. Method to extract antenna-launched hybrid polaritons. (a)** Raw data of the near-field image of a negative refraction lens, as recorded by our s-SNOM setup. The polaritons are launched by a gold antenna (length 3.2 µm; width 250 nm; thickness 50 nm). The α-MoO$_3$ is 240 nm thick. The Fermi energy of graphene is $E_F$= 0.5 eV. **(b)** Near-field image of antenna-launched hybrid polaritons corresponding to the raw data in (a), obtained by subtracting the vertical polariton fringes launched or reflected by the gold-SiO$_2$ interface. The scale bar indicates 3 µm. The horizontal orange and purple stripes on the upper part of the near-field images in (a and b) indicate the gold and SiO$_2$ substrates.

The spot size of the mid-IR beam under the AFM tip is about ~30 µm in lateral size, and therefore, it can cover a large area of the sample in Figure S21a, thus leading to both tip, antenna, and edge launching of polaritons. Indeed, different types of polariton fringes can be identified. The orange arrows indicate polaritons launched by the antenna. The vertical fringes indicated by the white arrows represent polaritons launched by the gold-SiO$_2$ interface. The black arrows indicate interference fringes of polaritons launched by the tip and reflected by the gold-SiO$_2$ interface. The raw data of the near-field image consists of 400×120 pixels (data points),

$$A_0 = \begin{bmatrix} a_1^1 & \cdots & a_1^{400} \\ \vdots & \ddots & \vdots \\ a_{120}^1 & \cdots & a_{120}^{400} \end{bmatrix},$$

where the polaritons launched by the antenna occupy the middle part of the image, while the area at the edge of the image only polaritons launched by the tip and the Au-SiO$_2$ interface do exist, which can be used as background signals,

$$A_1 = \frac{1}{10}\left( \begin{bmatrix} a_1^1 & \cdots & a_1^{400} \\ \vdots & \ddots & \vdots \\ a_5^1 & \cdots & a_5^{400} \end{bmatrix} + \begin{bmatrix} a_{116}^1 & \cdots & a_{116}^{400} \\ \vdots & \ddots & \vdots \\ a_{120}^1 & \cdots & a_{120}^{400} \end{bmatrix} \right).$$

$A_1$ is obtained from the average value of 10 rows of data from the upper and lower edges to achieve higher accuracy. Therefore, the signal associated with the polaritons launched by the antenna can be obtained from the difference B= $A_0$-$A_1$ (Figure S21b).

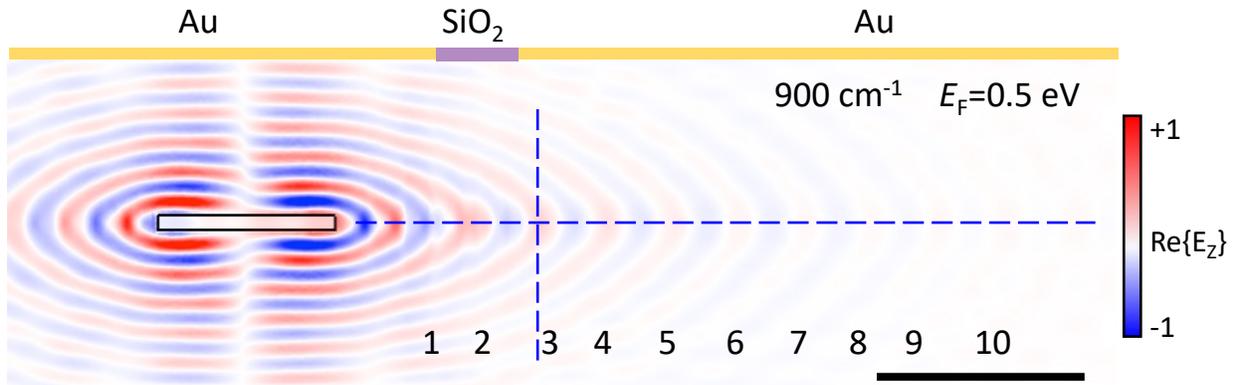

**Figure S22. Numerically simulated field distributions (Re {Ez}) compared with the measured results shown in Figure 4c.** The field is evaluated 50 nm above the surface of the heterostructure. The horizontal orange and purple stripes on top of the images represent areas with gold and SiO$_2$ substrate underneath, respectively. The scale bar indicates 3 μm.

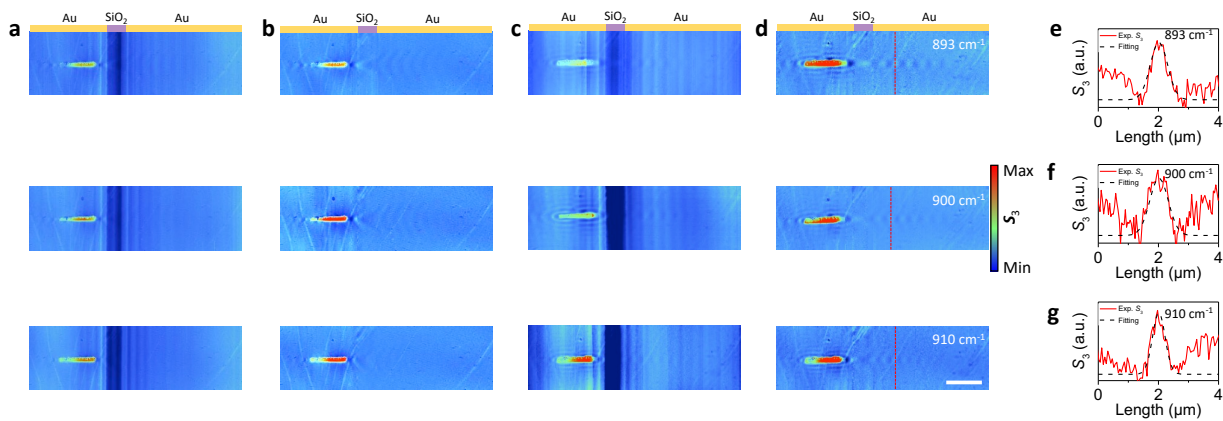

**Figure S23. Partial focusing of hybrid polaritons by a negative refraction lens for different illumination frequencies.** (**a, c**) Raw data of the near-field image without doping of graphene, recorded by our s-SNOM setup. (**b, d**) Near-field image of antenna-launched hybrid polaritons corresponding to the raw data in (a, c). The scale bar indicates 3 μm. The polaritons are launched by a gold antenna (length 3.2 μm; width 250 nm; thickness 50 nm). The α-MoO$_3$ is 240 nm thick. The Fermi energy of graphene is fixed at $E_F$= 0.5 eV in (c, d). The horizontal orange and purple stripes on the upper part of near-field images indicate the gold and SiO$_2$ substrates. (**e-g**) Near-field profiles taken at the positions marked by red vertical dashed lines in panels (d). The black dashed curves are Gaussian fittings.

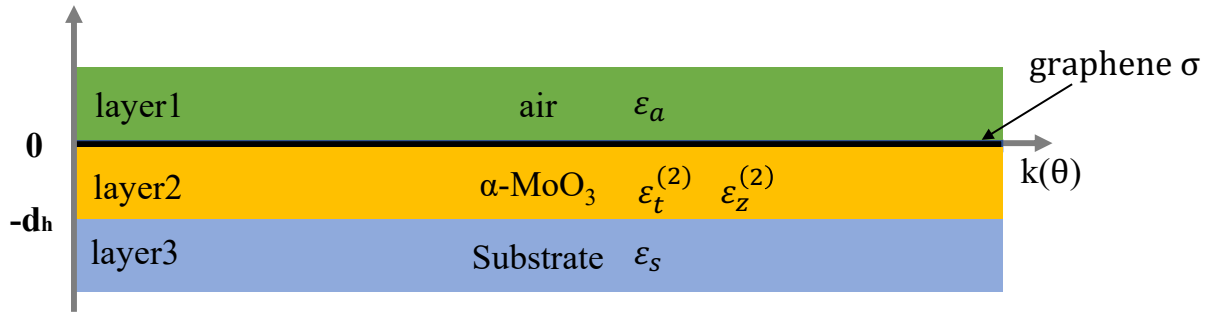

**Figure S24. Illustration of the geometry considered for the theoretical model.** Green, orange, and blue areas stand for different layers. Layer 1 ($z>0$, air) is a cover layer; Layer 2 ($0>z>-d_h$, graphene/α-MoO₃) is a middle layer; and Layer 3 ($z<-d_h$, SiO₂ or Au) is a substrate. Each layer is described as a dielectric material represented by its dielectric tensor. Monolayer graphene, located on the top of α-MoO₃ at $z=0$, is described as a zero-thickness current layer characterized by the surface graphene conductivity.

**References**

1 Zheng Z, *et al*. A mid-infrared biaxial hyperbolic van der Waals crystal. *Science advances*, **5**, eaav8690, (2019).

2 Kischkat, J. *et al*. Mid-infrared optical properties of thin films of aluminum oxide, titanium dioxide, silicon dioxide, aluminum nitride, and silicon nitride. *Applied optics*. Optical technology and biomedical optics **51**, 6789-6798, (2012).

3 Babar, S. & Weaver, J. H. Optical constants of Cu, Ag, and Au revisited. *Applied optics* **54**, 477, (2015).